\documentclass[10pt,journal,compsoc]{IEEEtran}
\ifCLASSOPTIONcompsoc
  \usepackage[nocompress]{cite}
\else
  \usepackage{cite}
\fi

\usepackage{xspace}
\usepackage{subfigure}
\usepackage{graphicx}
\usepackage{multirow}
\usepackage{hhline}
\usepackage{comment}
\usepackage{bigstrut}
\usepackage[dvipsnames]{xcolor}
\usepackage[sort&compress,square,comma,numbers]{natbib}
\usepackage[colorlinks=false, linkcolor=blue, citecolor=blue, bookmarks=true,pagebackref=false]{hyperref}
\usepackage{hyphenat}
\usepackage{float}
\usepackage{fancybox}
\usepackage[flushleft]{threeparttable}
\usepackage{array,booktabs}
\usepackage{amsmath}
\usepackage{outlines}
\usepackage[linesnumbered,ruled]{algorithm2e}
\usepackage[misc]{ifsym}
\usepackage{arydshln}
\usepackage{balance}
\usepackage{amsmath}
\usepackage{MnSymbol}
\usepackage{wasysym}
\usepackage{enumitem}
\usepackage{times}
\usepackage{fancyhdr}
\usepackage{courier}
\usepackage{booktabs}
\usepackage{pifont}
\usepackage{color,soul} 
\usepackage{tcolorbox}
\usepackage{ragged2e}
\usepackage[justification=centering]{caption}
\usepackage{natbib}
\usepackage{tikz}
\usepackage{capt-of}
\usepackage{makecell}
\newcommand\RQOne{Are SNA metrics more effective than code metrics for  Within-project SDP?}
\newcommand\RQTwo{Are SNA metrics more effective than code metrics for Cross-version SDP?}
\newcommand\RQThree{Are SNA metrics more effective than code metrics for Cross-project SDP?}

\newtcolorbox{mybox}[2][]{
top=0.15in,left=4pt,right=4pt,bottom=4pt,
fonttitle=\bfseries,
colbacktitle=gray,
colback=gray!5,
colframe=gray!40!black,
enhanced,
attach boxed title to top left={xshift=1.5em,yshift=-\tcboxedtitleheight/2},
boxed title style={size=small},
drop shadow={black!50!white},
title=#2,#1}

\ifCLASSINFOpdf
\else
\fi

\begin{document}
%
\title{Revisiting the Impact of Dependency Network Metrics on Software Defect Prediction}
%
%
%
%

\author{Lina~Gong,
        Gopi~Krishnan~Rajbahadur,
        ~Ahmed~E.~Hassan, and~Shujuan Jiang
\IEEEcompsocitemizethanks{\IEEEcompsocthanksitem Lina Gong is with the School of Computer Science and Technology, China University of Mining and Technology, China, and the School of Computer Science and Technology, Nanjing University of Aeronautics and
Astronautics, China. E-mail: gonglina@nuaa.edu.cn.

\IEEEcompsocthanksitem Gopi Krishnan Rajbahadur and Ahmed~E.~Hassan are with Software Analysis and Intelligence Lab (SAIL), School of Computing, Queen’s University, Canada. E-mail: {krishnan, ahmed}@cs.queensu.ca.
\IEEEcompsocthanksitem Shujuan Jiang is with with the School of Computer Science and Technology, China University of Mining and Technology, Xuzhou, China. E-mail: shjjiang@cumt.edu.cn.
\IEEEcompsocthanksitem Gopi~Krishnan~Rajbahadur is the corresponding author.
}

}

%
%


\IEEEtitleabstractindextext{%
\begin{abstract}
\justifying  
Software dependency network metrics extracted from the dependency graph of the software modules by the application of Social Network Analysis (SNA metrics) have been shown to improve the performance of the Software Defect prediction (SDP) models. However, the relative effectiveness of these SNA metrics over code metrics in improving the performance of the SDP models has been widely debated with no clear consensus. Furthermore, some of the common SDP scenarios like predicting the number of defects in a module (Defect-count) in Cross-version and Cross-project SDP contexts remain unexplored. Such lack of clear directive on the effectiveness of SNA metrics when compared to the widely used code metrics prevents us from potentially building better performing SDP models. Therefore, through a case study of 9 open source software projects across 30 versions, we study the relative effectiveness of SNA metrics when compared to code metrics across 3 commonly used SDP contexts (Within-project, Cross-version and Cross-project) and scenarios (Defect-count, Defect-classification (classifying if a module is defective) and Effort-aware (ranking the defective modules w.r.t to the involved effort)). We find the SNA metrics by themselves or along with code metrics improve the performance of SDP models over just using code metrics on 5 out of the 9 studied SDP scenarios (three SDP scenarios across three SDP contexts). However, we note that in some cases the improvements afforded by considering SNA metrics over or alongside code metrics might only be marginal, whereas in other cases the improvements could be potentially large. Based on these findings we suggest that the future work should: consider SNA metrics alongside code metrics in their SDP models; as well as consider Ego metrics and Global metrics, the two different types of the SNA metrics separately when training SDP models as they behave differently. 


\end{abstract}

}

\maketitle

\IEEEdisplaynontitleabstractindextext

%
\IEEEpeerreviewmaketitle

\IEEEraisesectionheading{\section{Introduction}\label{sec:introduction}}

\noindent \IEEEPARstart{A}{s} software scales and its complexity increases, more resources are required to test and inspect code. In practice, due to resource constraints such as time and cost, managers often allocate limited resources to test high-risk defective modules~\cite{Stefan2008tse, Shepperd2014tse, ghostra2015icse, yatish2019icse}. Thus, effectively identifying high-risk defective modules to be tested becomes particularly critical. To address this, Software Defect Prediction (SDP) models trained on software metrics are commonly used to identify high-risk defective modules~\cite{Ahmet2014ese, Shepperd2013tse, qin2011tse}. These SDP models are then used to effectively allocate testing resources. The use of software metrics (e.g., code metrics, SNA metrics, or process metrics) to train these SDP models play a crucial role in determining their performance. 




In 2008,~\citet{Thomas2008icse} demonstrated the effectiveness of using dependency network metrics along with source code metrics to improve the performance of SDP models. Such dependency network metrics (SNA metrics) were extracted from the dependency graph of the software modules by application of the Social Network Analysis. Zimmermann and Nagappan~\cite{Thomas2008icse} captured the dependency relations among software modules along 60 SNA metrics. They hypothesized that these SNA metrics captured the dependency structure of the software system better and thereby identified defects more effectively. They reported a 10\% improvement in the performance of the SDP model trained on SNA metrics for Windows Server 2003. 

\newcommand{\tabincell}[2]{\begin{tabular}{@{}#1@{}}#2\end{tabular}} 
\begin{table*}[tb]
\caption{Literature survey of all the prior studies that understand the effectiveness of SNA metrics over code metrics.}\label{table:survey}
\begin{threeparttable}
\begin{tabular}{l|r|r|r|r|r|r|r|r|r} 
\hline
\multirow{2}{*}{\textbf{Context}}& \multicolumn{3}{c|}{\textbf{Code metrics are more effective}} & \multicolumn{3}{c|}{\textbf{SNA metrics are more effective}} &\multicolumn{3}{c}{\textbf{SM metrics are more effective}}\\
\cline{2-10}
& \multicolumn{1}{c|}{\textbf{D-classification}} & \multicolumn{1}{c|}{\textbf{D-count}}& \multicolumn{1}{c|}{\textbf{E-aware}} &\multicolumn{1}{c|}{\textbf{D-classification}} & \multicolumn{1}{c|}{\textbf{D-count}}& \multicolumn{1}{c|}{\textbf{E-aware}} &\multicolumn{1}{c|}{\textbf{D-classification}} & \multicolumn{1}{c|}{\textbf{D-count}}& \multicolumn{1}{c}{\textbf{E-aware}}\\
\hline
\textbf{Within-project} &\tabincell{r}{ \cite{tosun2009promise} (small),\\ \cite{Satya2013apsec}\textsuperscript{*},\cite{Fang2014js}\textsuperscript{*}} & NAE & NAE &\tabincell{r}{\cite{Thomas2008icse}$^\Delta$,\\ \cite{Premraj2011esem}$^\Delta$}& NAE &\cite{Wan2016ist}$^\Delta$ & \tabincell{r}{\cite{Thomas2008icse}$^\Delta$,\\ \cite{Thanh2010icsm}, \cite{tosun2009promise} (Large),\\ \cite{Premraj2011esem}$^\Delta$, \cite{Satya2013apsec}\textsuperscript{*},\\ \cite{Fang2014js}\textsuperscript{*}} & \cite{Thomas2008icse}, \cite{Thanh2010icsm}& \cite{Thanh2010icsm}, \cite{Wan2016ist}$^\Delta$\\
\hline
\textbf{Cross-version} & \cite{Premraj2011esem}$^\circ$, \cite{Satya2013apsec} & NS & NAE & \cite{Premraj2011esem}$^\circ$ &  NS & \cite{Wan2016ist}$^\Delta$ & \cite{Premraj2011esem}$^\circ$ & NS & \cite{Wan2016ist}$^\Delta$\\
\hline
\textbf{Cross-project} & \cite{Premraj2011esem}$^\bullet$, \cite{Satya2013apsec} & NS & \cite{Wan2016ist}$^\bullet$ & \cite{Premraj2011esem}$^\bullet$ & NS & \cite{Wan2016ist}$^\bullet$ &NAE & NS & NAE\\
\hline
\end{tabular}
 \begin{tablenotes}
 \scriptsize
 \item 1. Simple Combination (SM) metrics- Code and SNA metrics simply combined
 \item 2. \textbf{D-Classification}- Defect-classification; \textbf{D-count}- Defect-count; \textbf{E-aware}- Effort-aware
 \item 3. \textbf{Small:} Large performance impact was only found on small software projects
 \item 4. \textbf{Large:} Large performance impact was only found on large software projects
 \item 5. NAE - Not As Effective; NS- Not studied
 \item 6. *: code metrics and SM metrics were found to be equally effective; $\Delta$: SNA metrics and SM metrics were found to be equally effective; $\bullet$: code metrics and SNA metrics were found to be equally effective; $\circ$: All three metrics were found to be equally effective (i.e., no difference was observed in effectiveness of the studied metrics).
\end{tablenotes}
\end{threeparttable}
\end{table*}

Since the early work of~\citet{Thomas2008icse}, several studies have explored the effectiveness of SNA metrics over code metrics for SDP and have come up with conflicting conclusions. Such prior studies examined the effectiveness of SNA metrics in various SDP contexts (i.e., Within-project, Cross-version and Cross-project) and scenarios (Defect-count (predicting the number of defects)), Defect-classification (classifying if a module is defective) and Effort-aware (ranking of defect prone modules w.r.t required effort)). The reached conclusions can be grouped into three categories: 1) studies that argue SNA metrics along with code metrics can indeed improve the performance of SDP models~\cite{Thomas2008icse,Thanh2010icsm}, 2) studies that argue SNA metrics provide no significant advantage over using just the code metrics~\cite{Satya2013apsec}, 3) finally, studies that argue SNA metrics are effective for a limited number of SDP contexts and scenarios~\cite{tosun2009promise, Wan2016ist, Premraj2011esem}. 

For instance,~\citet{Thanh2010icsm} showed that SNA metrics alongside code metrics indeed help in the improving the performance of SDP models in the Within-project context. Whereas,~\citet{Satya2013apsec} found that SNA metrics offered no significant advantage over code metrics in any of the SDP contexts. Different from both of these studies,~\citet{Wan2016ist} reported that SNA metrics offered a slight improvement over code metrics only in the Within-project and Cross-version SDP contexts. Such conflicting results prevents us from recommending the most appropriate metrics for building effective models. Table~\ref{table:survey} presents a survey of prior SDP studies that explore SNA metrics (More details about the survey in Section~\ref{sec:related_works}).


 

We further note that, \textit{\textbf{none}} of the prior studies attempt to rationalize the varying effectiveness of SNA metrics for a given SDP context and/or scenario. Hence our key contributions are:

\begin{enumerate}[wide = 0pt, itemsep = 3pt]
\item We examine the effectiveness of SNA metrics across 3 common SDP contexts and scenarios on nine projects across 30 versions.
\item We attempt to rationalize the varying effectiveness of SNA metrics across the studied SDP contexts and scenarios.
\item In contrast to the prior studies, we study the effectiveness of separately considering the different types of SNA metrics (i.e., Global metrics (GN) and Ego metrics (EN)) for training a SDP model. In addition, we also study the effectiveness of using a simple filtering strategy to combine the code and SNA metrics (COM metrics) over simply combining them (SM metrics).

\end{enumerate}

In our study, we focus only on comparing the effectiveness of SNA metrics over code metrics (and not other metric families like process metrics), as the history of a software project is not always readily available. In cases such as those, only metrics based on source code could be used to build models. We structure our study along the following research questions:

\begin{itemize}[wide = 0pt, itemsep = 3pt]
\item \textbf{RQ1: \RQOne}\\
Models trained on SM and COM metrics outperform models trained on all other metric families across Defect-Count and Defect Classification scenarios. In addition, models trained on SNA metrics outperform models trained on other metric families for Effort-aware scenarios. We observe that, such a result is because SNA metrics add additional information to the code metrics. Therefore, we recommend researchers and practitioners to consider the SNA metrics alongside code metrics in the Within-project context.



\item \textbf{RQ2: \RQTwo}\\
The results are varied for the Cross-version context across the studied scenarios. Models trained on code metrics outperform models trained on other metric families for the Defect-Count scenario. For the Defect-classification scenario, models trained on SNA metrics and code metrics perform similarly. Models trained on EN and COM metrics outperform models trained on other metric families for the Effort-aware scenarios. In addition, though code metrics are effective for Defect-count scenario, we observe that the models trained on GN metrics have better performance in identifying the number of defects on modules whose defectiveness changes across versions. These results indicate that considering SNA metrics alongside code metrics might be useful. In addition, these results highlight the need for considering GN and EN metrics separately. 

\item \textbf{RQ3: \RQThree}\\
Models trained on code metrics outperform models trained on other metric families for the Defect-Count scenario. For the Defect-classification scenario, models trained on SNA metrics (GN, SM and COM) metrics have similar performance as the models trained on code metrics on 3 of the 5 studied performance measures. Finally, the models trained on GN metrics outperform models trained on other metric families for the Effort-aware scenarios.
\end{itemize}
From the results we observe that considering SNA metrics by themselves or alongside code metrics is effective across all the SDP contexts and 5 out of 9 studied SDP scenarios. However, we note that in some cases the improvements afforded by considering SNA metrics over or alongside code metrics might only be marginal, whereas in other cases the improvements could be potentially large. In light of these findings, we recommend researchers and practitioners to always consider including SNA metrics to train their models, particularly given the ease of extracting these metrics. In addition, to foster open science and to ensure that our results are replicable we provide a replication package for our study~\footnote{https://github.com/glnmzx888/SNA}. 

 \noindent{\textbf{Paper organization.} Section~\ref{sec:related_works} motivates our study and discusses related work. Section~\ref{sec:background} explains the metrics, models and evaluation measures used in our study. Section~\ref{sec:study_design} presents our study design, including studied projects and experiment setup. Section~\ref{sec:case_study_results} presents and discusses our experimental results. Section~\ref{sec:implications} presents the implications that can be inferred from our results. Section~\ref{sec:threats_to_validity} discusses the threats to the validity of our observations. Finally, we conclude our study in section~\ref{sec:conclusions}.}
\section{Motivation and Related work}\label{sec:related_works}
\subsection{Motivation}
SDP has become a very active research field~\cite{kamei2012tse,Hall2012tse,Assar2016ese}. 
Researchers have proposed many metrics based on various aspects to build models, such as code metrics~\cite{Basili1996tse, Briand2000jss,Ostrand2004sigsof, Nagappan2006icse, zhou2010jss} and process metrics~\cite{Foyzur2013icse, Madeyski2014sqj, Nagappan2006sre}. Code metrics leverage the source code of the software such as Halstead~\cite{Halstead1977elsevier} and McCabe~\cite{McCabe1976tse} metrics and Chidamber-Kemere (CK)~\cite{Chidamber1994tse} metrics. Process metrics leverage the software history such as change complexity~\cite{Ahmed2009icse}, change coupling~\cite{Marco2009re}, change history~\cite{graves2000tses}, and developer metrics~\cite{meneely2008sigsoft, tian2013ase, bird2011sigsoft}.

When the history of a software is not readily available, only metrics based on the source code can be used to build models. For instance, consider small companies or start ups where there is no established process to capture the software evolution~\cite{Eriks2019ese}. In addition, companies might not always be willing to share history based metrics due to privacy concerns~\cite{peter2013tse, peters2015icse}. Furthermore, as~\citet{Nagappan2010sre} argues process metrics are presumably more expensive to collect. Therefore, it is important to be able to leverage the information available in the source code of a software for SDP. 


There exist many rich dependencies between the software modules in addition to the static details about the source code such as lines of code. For instance, data and call dependencies could be extracted from the source code. Several prior studies make use of such dependencies to build models~\cite{shin2009MSR, shin2012ese,adrain2006isese,Thomas2008icse}. Most notably,~\citet{Thomas2008icse} demonstrated the effectiveness of dependency network metrics that were extracted using SNA on the dependency graph of the software modules. Such SNA metrics have been the most widely adopted set of source code based dependency metrics in SDP. However, as we mention earlier, the effectiveness of SNA metrics to train models is inconclusive. To capture the common consensus on the effectiveness of SNA metrics, we conducted a thorough literature survey of prior SDP studies that examined the effectiveness of SNA metrics over code metrics. We conducted the literature survey following the process that was outlined by~\citet{wohlin2012experimentation}. We first searched google scholar\footnote{https://scholar.google.ca/} with a single query for the terms ``defect prediction" and ``Social Network Analysis" then we went through the abstract of all the returned papers. We shortlisted the papers that compared SNA metrics to code metrics. Following which, we conducted a forward and reverse snowball search~\cite{wohlin2012experimentation} for all the papers to which a given paper refers. We then went through each of them to locate any other relevant studies. We continued the forward and reverse snowball search for each of the collected papers recursively until we exhaust all the relevant papers. We finally arrived at 8 studies. We present the results of our literature survey in Table~\ref{table:survey}. We present and categorize each study based on the SDP context and scenario under which the effectiveness of the SNA metrics were studied. Each row of the Table~\ref{table:survey} captures the SDP context under which a given study compares the effectiveness between SNA and code metrics. Each column of the Table~\ref{table:survey} captures the metric family that a given study finds more effective. Additionally, each of these columns are further sub-divided 3 columns, which capture the SDP scenario under which a given study observed the metric family in this column to be more effective. If a given study found two metric families to be equally effective, we then present the study in both the columns and mark it with a symbol to present which two metric families were found to be equally effective. For instance, consider the study by~\citet{Premraj2011esem}. They found that for the Defect-classification scenario, in the Cross-version context, models trained on code, SNA and SM (code and SNA metrics) metrics perform similarly. Hence, we represent that in our Table~\ref{table:survey} by having an entry for the study~\cite{Premraj2011esem} on all the three columns pertaining to each metric family and SDP scenario. Furthermore, we mark the study with a $``\circ"$ symbol represent the same.



From the results presented in Table~\ref{table:survey} we observe four shortcomings about prior studies of the effectiveness of SNA metrics. 

\smallskip\noindent\textbf{First, there is no clear consensus on whether code metrics consistently outperform SNA metrics or the SM metrics.} For instance, consider the Cross-version context.~\citet{Satya2013apsec} find that code metrics outperform SNA metrics for the Defect-classification scenario. Whereas,~\citet{Premraj2011esem} find that code, SNA metrics perform similarly for the same scenario across different classifiers. Such conflicts in the results could be observed across all the SDP context and scenarios in Table~\ref{table:survey}. 

\smallskip\noindent\textbf{Second, prior studies mostly limit themselves to one or two SDP scenarios.} For instance, consider the studies by~\citet{tosun2009promise} and~\citet{song2019tse} which consider only the within-project context and the Defect-classification scenario. Though some studies like~\citet{Premraj2011esem} and~\citet{Wan2016ist} span all the three common SDP contexts, they limit themselves to a single SDP scenario. For instance,~\citet{Premraj2011esem} focus on the Defect-classification scenario, while~\citet{Wan2016ist} focus on the Effort-aware scenario. Moreover, both the studies use different datasets and models making it impossible to reach a common finding that applies across the SDP scenarios.

\smallskip\noindent\textbf{Third, the effectiveness of SNA metrics in SDP scenarios like Defect-count scenario in the Cross-version and Cross-project contexts remain unexplored.} Table~\ref{table:survey} denotes the SDP scenarios under which the effectiveness of SNA metrics remain unexplored. Lack of research in such SDP scenarios affects our ability to recommend SNA metrics widely.

\smallskip\noindent\textbf{Finally, none of the prior studies explore the effectiveness of different types of SNA metrics (GN and EN metrics).} The GN metrics capture the dependency characteristics of a given module in the context of the whole software system. Whereas, EN metrics capture the dependency characteristics of a given software module with respect to its neighboring nodes~\cite{Thomas2008icse} (see Section~\ref{sec:SNA} for more details about the metrics). Despite the differences between GN and EN metrics, none of the prior studies outlined in Table~\ref{table:survey} consider these separately. Therefore, in our study, we consider them separately and together to understand their effectiveness in relation to code metrics for building models. 




\subsection{Related work}

\smallskip\noindent\textbf{Studies comparing various software metrics:} Prior studies have compared the effectiveness of various metric families for SDP.~\citet{Raimund2008icse} demonstrated that process metrics are more effective than code metrics for SDP.~\citet{Ahmed2009icse} proposed an entropy based metric and examined its effectiveness relative to the number of prior faults and prior modifications.~\citet{Arisholm2010jss} evaluated the effectiveness of code and process metrics for SDP and observed that code metrics outperformed for the Defect-classification scenario, but under performed for Effort-aware scenario. Whereas, process metrics were effective in the Effort-aware scenario. Alternatively,~\citet{Nagappan2010sre} found that change bursts were more effective than complexity metrics, code churn, and organizational structure. However,~\citet{Foyzur2013icse} later observed that process metrics were more effective than code metrics. 

Different from these studies, we study the effectiveness of SNA metrics over code metrics and investigate the rationale for the effectiveness of SNA metrics. 

\smallskip\noindent\textbf{Studies of other source code based dependency metrics:} Different from the SNA metrics computed on the dependency graph, other dependency metrics have been used for SDP.~\citet{adrain2006isese} leveraged the import relationships between software components for SDP and showed that the imported software components could predict post-version defects more effectively than random guesses.~\citet{Turhan2008seaa} proposed a static call graph based ranking (CGBR) framework, which was based on both static code metrics and the call dependency relations between software modules. Shin et al.~\cite{shin2009MSR, shin2012ese} investigated the effectiveness of the calling structure on SDP.

\section{Background}\label{sec:background}
In this section, we briefly describe the software metrics, SDP contexts and SDP scenarios considered in our study.
\subsection{Software metrics}\label{sec:metrics}
Different software metrics capture details about different facets of a software project. We briefly describe the six different software metrics that we use in our study: Code metrics, SNA metrics, EN metrics, GN metrics, Simple Combination (SM) metrics and COM metrics.

\subsubsection{Code metrics} 
Code metrics quantify various properties of the source code of a software project. For instance, size of the project is typically quantified by the lines of code metric. These code metrics can be computed at various granularities: method-level, class-level and file-level. 

\smallskip\noindent\textit{Method-level metrics:} These metrics quantify the various properties of methods. We use the Halstead~\cite{Halstead1977elsevier} and McCabe~\cite{McCabe1976tse} metrics to study method level properties as they are some of the most popular method-level metrics in SDP.

\smallskip\noindent\textit{Class-level metrics:} These metrics are observed at the class level and are only used for object-oriented (OO) projects. We use the Chidamber-Kemere (CK)~\cite{Chidamber1994tse} metrics (e.g., Coupling Between Object (CBO), Depth Of Inheritance Tree (DIT)) as the class-level metrics. 

\smallskip\noindent\textit{File-level metrics:} File-level metrics are metrics derived from each file. They also include metrics that are generated by aggregating method-level or class-level metrics. For instances, we aggregate the Method-level metrics by max, min and mean at the file level. 

In our study, we consider the metrics at all the three granularties above mentioned as \textbf{code metrics}. The 54 code metrics used in our study are listed in Table~\ref{table:metric}. For a detailed description about each of the studied code metrics, we refer the readers to \texttt{SciTools} website\footnote{\url{https://scitools.com/feature/metrics/}}. 

\begin{table*}[!htb]
\centering
\caption{Software metrics used in our study}\label{table:metric}
\hspace{0.5cm}
\centering
 \begin{tabular} {l|l|p{30pt}|p{300pt}|r} 
\hline
\textbf{Metric group}&\textbf{Metrics}&\textbf{Level} & \textbf{Metrics} & \textbf{Count}\\
\hline
\multirow{3}{*}{\textbf{Code metrics}}&\multirow{3}{*}{\textbf{Code}}&\textbf{File} & AvgCyclomatic, AvgCyclomaticModified, AvgCyclomaticStrict, AvgEssential, AvgLine, AvgLineBlank, AvgLineCode, Avg-LineComment, CountDeclClass, CountDeclClassMethod, CountDeclClassVariable, CountDeclFunction, CountDeclInstanceMethod, CountDeclInstanceVariable, CountDeclMethod, CountDeclMethodDefault, CountDeclMethodPrivate, CountDeclMethodProtected, CountDeclMethodPublic, CountLine, CountLineBlank, CountLineCode, CountLineCodeDecl, CountLineCodeExe, CountLineComment,
CountSemicolon, CountStmt, CountStmtDecl, CountStmtExe, MaxCyclomatic, MaxCyclomaticModified, MaxCyclomaticStrict, RatioCommentToCode, SumCyclomatic, SumCyclomaticModified, SumCyclomaticStrict, SumEssential &37\\
\cline{3-5}
&&\textbf{Class} & CountClassBase, CountClassCoupled, CountClassDerived, MaxInheritanceTree, PercentLackOfCohesion & 5\\
\cline{3-5}
&&\textbf{Method} & CountInput \{Min, Mean, Max\}, CountOutput \{Min, Mean, Max\}, CountPath \{Min, Mean, Max\}, MaxNesting \{Min, Mean, Max\} & 12\\
\hline
\multirow{3}{*}{\textbf{SNA metrics}}&\textbf{\textbf{EN}}& \textbf{File} &Size \{in, out, un\}, Ties \{in, out, un\}, Pairs \{in, out, un\}, Density \{in, out, un\}, WeakComp \{in, out, un\}, nWeakComp \{in, out, un\}, 2StepReach \{in, out, un\}, ReachEfficency \{in, out, un\}, Brokerage \{in, out, un\}, nBrokerage \{in, out, un\}, EgoBetween \{in, out, un\}, nEgoBetween \{in, out, un\}, EffSize(ego), Efficiency(ego), Constraint(ego), Hierarchy(ego) & 40\\
\cline{2-5}
&\textbf{GN}&\textbf{File} & Degree, Eigenvector, power, Closeness, Betweenness, Information, dwReach, EffSize, Efficiency, Constraint, Hierarchy & 24\\
\cline{2-5}
&\textbf{SNA}& \textbf{File}&EN+GN&64\\
\hline
\multirow{2}{*}{\textbf{Combined Metric}}&\textbf{SM}&\textbf{File}&Code+SNA&118\\
\cline{2-5}
&\textbf{COM}&\textbf{File}& CountDeclMethodDefault, AvgLineComment, CountDeclMethodProtected, RatioCommentToCode, AvgLineBlank, MaxInheritanceTree, CountClassDerived, CountInput\{Min\}, CountOutput\{Max\}, CountOutput\{Min\}, MaxNesting\{Min\}, Densit(in), pWeakC(in), nBroke(in), nEgoBe(in), pWeakC(un), nEgoBe(un)&17\\
\hline
\end{tabular}
\end{table*}

\subsubsection{SNA metrics} \label{sec:SNA}

SNA metrics are extracted by applying Social Network Analysis on the dependency network of software project~\cite{Thomas2008icse}. Dependency networks are graphs whose nodes are software modules and edges are dependencies between these modules. There are two main families of dependency relationships: data dependency and call dependency. In such a dependency network that includes both call and data dependencies, each node belongs to two types of networks: ego network and global network. 

\smallskip\noindent\textbf{Ego network metrics (EN).} The Ego network of a node consists of the node itself and its neighboring nodes. For each node, there exists three types of ego networks: In, Out and Undirected. (In) ego of a node only considers itself and nodes which it depends on. (Out) ego of a node only considers itself and nodes that depend on it. (Undirected) ego of a node considers a combination of In and Out. In our study, we compute all three types of ego networks.

\smallskip\noindent\textbf{Global network metrics (GN).} Global network refers to the entire dependency graph, which can measure the importance of nodes in the context of the entire software system. In our experiments, we compute structural holes and centrality as global network (GN) metrics.

\smallskip\noindent\textbf{SNA metrics (SNA).} Combination of EN and GN metrics together makes up the SNA metrics.

Table~\ref{table:metric} describes the EN and GN metrics. We consider EN, GN and SNA metrics as three different metric families in our study. We present more details about each of the studied SNA metric in Table 9 of Appendix~A.

\subsubsection{Combined Metrics}

We consider two different ways of combining code metrics and SNA metrics to observe if a combination of these metrics outperform the individual metrics.

\smallskip\noindent\textbf{Simple combination metrics (SM).} SM metrics are a simple combination of both code and SNA metrics. Code and SNA metrics are used without any metric being discarded. Such a way of combining the metrics has been used by all the prior studies listed in Table~\ref{table:survey}.

\smallskip\noindent\textbf{Common metrics (COM).} We take SM metrics and use a simple filtering strategy to filter metrics. We do so as, among the combined metrics, the stronger signal contained by some metrics could potentially be masked by the other collinear metrics. Therefore, we employ a Variance Inflation Factor (VIF) filter that we describe below to filter and combine the metrics.

\noindent\textit{Variance Inflation Factor (VIF) filter:} We use the Variance Inflation Factor (VIF) to remove multicollinearity among the studied metrics. The VIF score quantifies the collinearity of a metric by computing how well the given metric can be represented by the other studied metrics~\cite{Jirayus2018ICMSE, Fabio2019tse}. The VIF score of a metric is computed by first building a linear regression model using the other studied metrics to predict the metric under examination. Then the \textbf{R\textsuperscript{2}} score~\cite{cameron97joe} of the built regression model is computed. Finally, the VIF score of the metric under examination is given as $\frac{1}{1-R\textsuperscript{2}}$. 

A large VIF score for a given metric indicates that the given metric is highly collinear with the other studied metrics. Therefore, for each project, we only consider the metrics whose VIF score is less than 10. We use a threshold of 10 since several prior studies~\cite{bettenburg2010icpc, Wan2016ist} consider the metrics that have a VIF score $>$10 to be highly collinear with the other metrics.


The method of filtering metrics varies based on the context. In the Within-project context, we simply remove all the metrics in a given project whose metrics have a VIF$>$10. In the Cross-version context, we define COM metrics as those SM metrics whose VIF$<$10 in all versions of a project. In the Cross-project context, we use the COM metrics whose VIF$<$10 across all the studied projects. We report the 17 metrics that are present in all projects and have a VIF$<$10 in Table~\ref{table:metric}.

\subsection{Software Defect Prediction (SDP) models}

In this sub-section, we detail the common SDP contexts and scenarios that we study.

\subsubsection{SDP scenarios}
Models are typically constructed for the following three SDP scenarios: Defect-count, Defect-classification, and Effort-aware scenarios.

\smallskip\noindent\textbf{Defect-count scenario} is where a model (Defect-count model) predicts the number of defects in a given software module~\cite{Andreou2016jss,Rathore2017eswa,Rathore201sc}. 

\smallskip\noindent\textbf{Defect-classification scenario} is where a model (Defect-classification model) classifies whether a given software module is defective or not~\cite{Menzies2007tse, Hall2012tse}.

\smallskip\noindent\textbf{Effort-aware scenario} is where a model (Effort-aware model) ranks software modules based on their probability of being defective with respect the effort required to identify the defects~\cite{marco2012ese,kamei2012tse,Assar2016ese,Gupta2017cp}. For instance, ranking the software modules based on their probability of being defective and the lines of code that one needs to inspect in order to identify the defects.

\subsubsection{SDP contexts}\label{sec:sdp_contexts}
SDP contexts can be divided into: Within-project context, Cross-version context and Cross-project context. 

\smallskip\noindent\textbf{Within-project context} is where the training data and testing data are derived from the same version of a project~\cite{Tantithamthavorn2016icse,tantithamthavorn2017tse}. Commonly, a part of the data from a version is used for training a model and the remaining is used for testing the trained model.

\smallskip\noindent\textbf{Cross-version context} is where the training data to predict the defects in the current version of a project is comprised of data from its prior versions~\cite{xu2018icpc,xu2018jss}. For instance, in the Cross-version context, one uses versions V\textsubscript{1},V\textsubscript{2},\dots,V\textsubscript{n-1} to train a model to predict defects in the version V\textsubscript{n}.

\smallskip\noindent\textbf{Cross-project context} is where the training data to predict the defects in the current version of a project is comprised of data from completely different projects~\cite{Hosseini2019tse,xia2016tse}. For instance, to predict the defects in the version V\textsubscript{k} of a project P\textsubscript{B}, the model in the Cross-project context is typically trained on the latest version V\textsubscript{m} belonging to project P\textsubscript{A}.

\subsection{Model evaluation measures}\label{sec:measures}
In this sub-section, we detail the different performance measures that we use for evaluating our trained models.
\subsubsection{Performance measures for Defect-count models}
To evaluate the effectiveness of Defect-count models, we use four different performance measures: Mean Absolute Error, Adjusted R square, Root mean squared error and Spearman correlation coefficient~\cite{Thanh2010icsm, William2016ist}. We use these residual based performance measures over the commonly used performance measures like Magnitude of Relative Error (MRE) or Pred(X). We do so as several prior studies point out that MRE and Pred(X) are biased~\cite{Martin2012ist, Ekrem2013tse, ingunn2005tse, kitchenham2001ips}. In particular,~\citet{ingunn2005tse} showed that the MRE measure yields completely different values for two prediction models even when their residuals are similar. Due to these shortcomings,~\citet{ingunn2005tse} recommended the use of residual based performance measures (similar to the ones that we use) to evaluate the performance of regression models. We briefly explain these measures below. 

\smallskip\noindent\textbf{Mean Absolute Error (MAE).} MAE computes the mean of the absolute difference between the number of predicted defects by the Defect-count model and actual number of defects for each module~\cite{yatish2019icse}. The lower the MAE value, the better the effectiveness of the Defect-count model.

\smallskip\noindent\textbf{Root Mean Squared Error (RMSE).} RMSE measures the square root of the average of squared differences between the number of predicted defects by a Defect-count model and actual number of defects for each module~\cite{chai2014gmd}. A lower RMSE score indicates better performance.

\smallskip\noindent\textbf{Adjusted R Square (Adj R\textsuperscript{2}).} Adj R\textsuperscript{2} measures the amount of variance of the dependent variable that the Defect-count model captures with respect to the number of used metrics~\cite{Thanh2010icsm}. The Adj R\textsuperscript{2} indicates how good a defect count model fits the training data. A higher Adj R\textsuperscript{2} value indicates a better performing model.

\smallskip\noindent\textbf{Spearman Correlation Coefficient (Spearman).} Spearman correlation coefficient indicates the correlation between predicted number of defects and actual number of defects~\cite{Thanh2010icsm}. A high spearman values indicates that a Defect-count model can accurately predict the number of defects.


\subsubsection{Performance measures for Defect-classification models}\label{sec:classification_measure}

To evaluate the effectiveness of Defect-classification models, we use the following performance measures such as: Area Under the Reciever-operator Characteristic Curve (AUC), Matthews correlation coefficient (MCC), Precision, Recall and Brier score similar to several prior studies~\cite{ghostra2015icse, Davidebmc2020, Thomas2008icse,rajbahadur2019impact,tantithamthavorn2019tse}. A high value for AUC, Precision, Recall and MCC indicates a good performing model. A more detailed explanation about each of the studied performance measure is provided in the Table 10 of Appendix~B.



\subsubsection{Performance measures for Effort-aware models}

To evaluate the effectiveness of Effort-aware models, we use the cost-effectiveness (CE) curve~\cite{Arisholm2010jss,Wan2016ist} and effort reduction (ER) measures~\cite{Yonghee2011tse,Wan2016ist}. 

\begin{figure}[!htbp]
\centering
\includegraphics[width=\columnwidth]{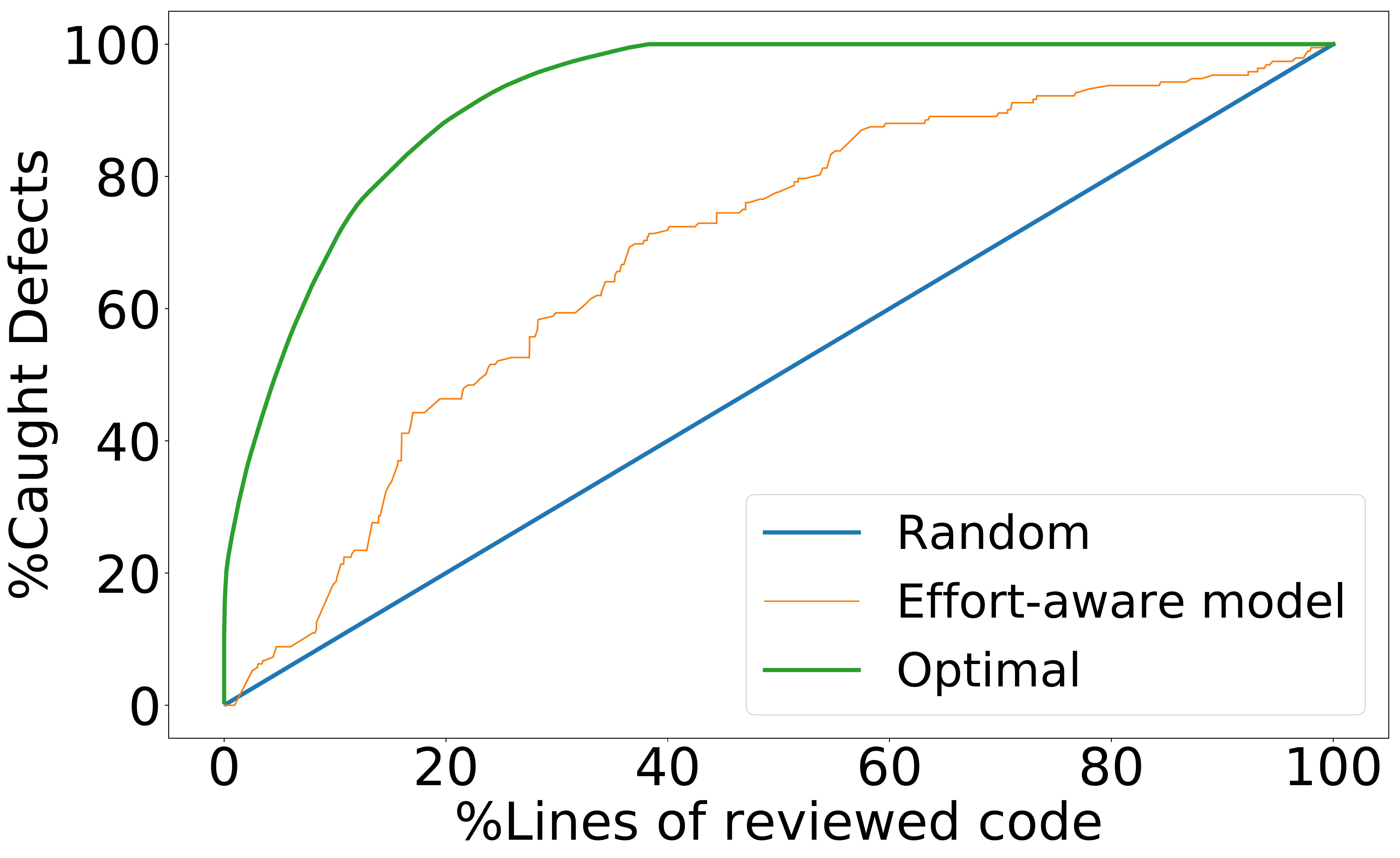}
\caption{Cost-Effectiveness curve for an Effort-aware model}\label{fig:CE_curve}
\end{figure}

\smallskip\noindent\textbf{Area under the Cost-Effectiveness Curve (CE).} The area under the cost-effectiveness curve quantifies the required effort to identify defects. It does so by computing the area under the Cost-Effectiveness (CE) curve. Figure~\ref{fig:CE_curve} shows a CE curve for an Effort-aware model. The x-axis of the Figure~\ref{fig:CE_curve} is the percentage of reviewed LOC and y-axis is the percentage of discovered defects. If the majority of defects can be identified by reviewing a small percentage of the LOC then the model performance is optimal. Such an optimal scenario is depicted in Figure~\ref{fig:CE_curve} with the green Optimal line. Contrastingly, the worst CE curve is one in which the Effort-aware model's ranking is completely random and the percentage of found defects linearly increases with the percentage of reviewed LOC. Such a scenario is depicted in Figure~\ref{fig:CE_curve} with a blue line. The orange line indicates how many defects could be caught with respect to the percentage of LOC that need to be reviewed for the prediction by the Effort-aware model. The closer the orange line is to the green Optimal line, the better. The area under the orange line is given as the Area under the CE curve. We calculate the Area under the CE curve with Formula~\ref{eq:ce} as suggested by~\citet{Arisholm2010jss}. A high CE value indicates that the Effort-aware model can identify most of the defects by investigating the least possible LOC. $\pi$ is a cut-off value which shows the percentage of LOC at which the Area under the CE curve is calculated. In our experiments, we use $\pi=1$ to report the CE values to present a holistic summary.



\begin{equation}
\label{eq:ce}
CE_\pi=\frac{Area_\pi(m)-Area_\pi(Random)}{Area_\pi(Optimal)-Area_\pi(Random)}
\end{equation}

\smallskip\noindent\textbf{Effort Reduction (ER)}
ER indicates the ratio of the reduced LOC to be reviewed when using an Effort-aware model to LOC to be reviewed when using a random model to achieve the same recall as the Effort-aware model. We use the ER measure as outlined by~\citet{Yonghee2011tse} and~\citet{Wan2016ist}. We calculate the ER for our Effort-aware model (EM) with the following Formula~\ref{eq:er} . 

\begin{equation}
\label{eq:er}
ER(EM)=\frac{Effort(Random)-Effort(EM)}{Effort(Random)}
\end{equation}

In Formula~\ref{eq:er},~\textit{Effort(EM)} is defined as the ratio of LOC to be inspected (i.e., where the Effort-aware model predicts the module to be defective) to the total LOC. We calculate the \textit{Effort(EM)} with Formula~\ref{eq:effort}. In Formula~\ref{eq:effort}, p\textsubscript{i} is 1 if the Effort-aware model considers the module defective and 0 otherwise. Similarly, to compute the effort for the random model~\textit{Effort(Random)}, we use Formula~\ref{eq:effort_R}, where d\textsubscript{i} is the number of defects in the software module i. Finally, we compute the ER with the Formula~\ref{eq:er}. A high ER values indicates that the Effort-aware model performs well~\cite{Wan2016ist}.


\begin{equation}
\label{eq:effort}
Effort(EM)=\frac{\sum_{i=1}^{n}l_i \times p_i}{\sum_{i=1}^{n}l_i}
\end{equation}

\begin{equation}
\label{eq:effort_R}
Effort(Random)=\frac{\sum_{i=1}^{n}d_i \times p_i}{\sum_{i=1}^{n}d_i}
\end{equation}


\section{Study design}\label{sec:study_design}
\begin{figure*}[!tb]
\centering
\includegraphics[width=5.5in]{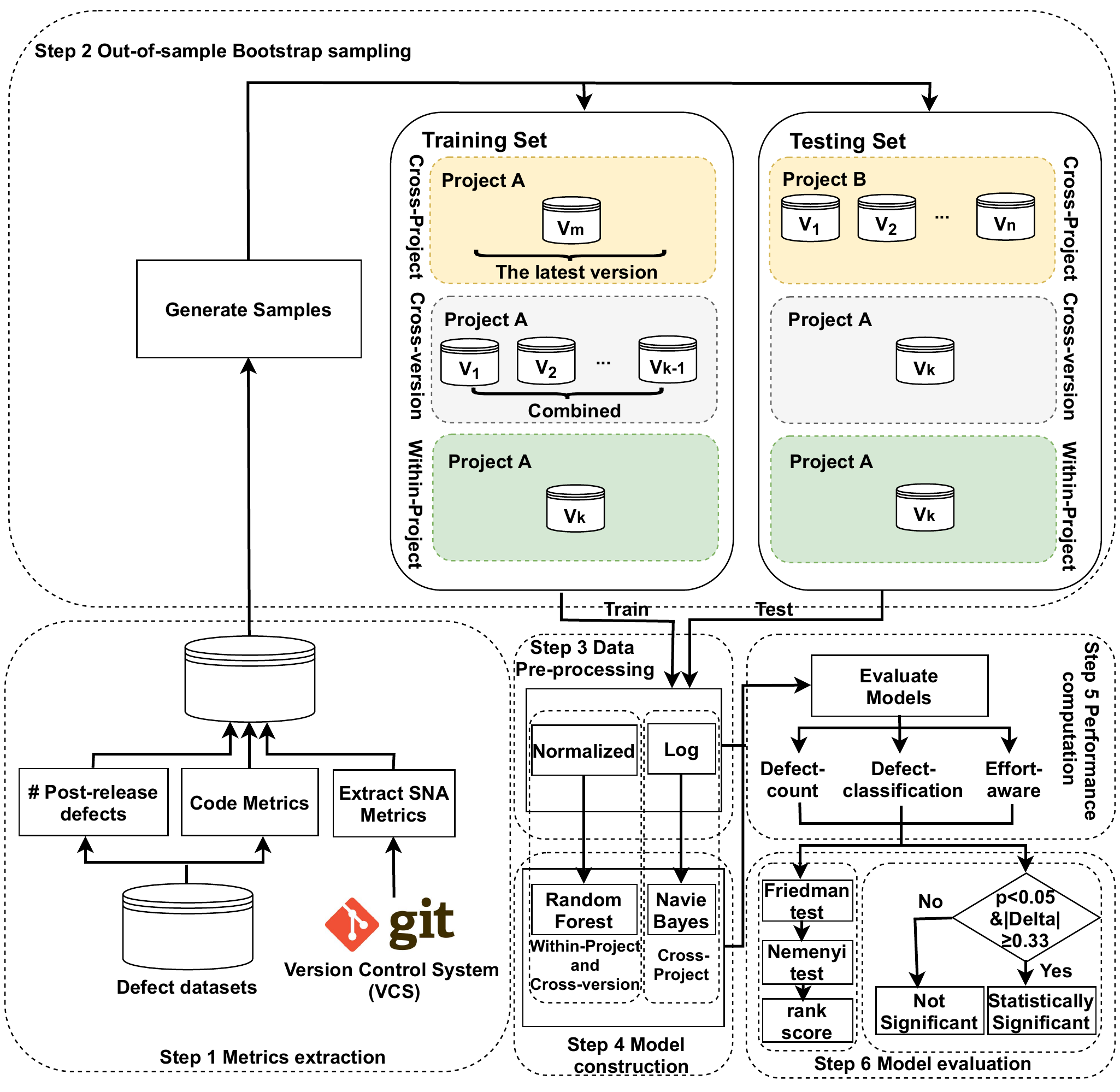}
\caption{Overview of our study design}\label{fig:overview}
\end{figure*}

In this section, we describe our study design to answer our research questions. An overview of our study design is shown in Figure~\ref{fig:overview}.

\subsection{Studied Projects}
We conduct our experiments on the same software projects as~\citet{yatish2019icse}. These projects represent a carefully curated list of benchmark defect datasets, which consists of 32 versions of 9 open-source software projects including ActiveMQ, Camel, Derby, Groovy, HBase, Hive, JRuby, Lucene and Wicket. Table~\ref{table:dataset} provides a detailed overview of our studied projects. The column ``\#defects” in Table~\ref{table:dataset} represents the total number of post-release defects that were identified each version.~\citet{yatish2019icse} extracted the post-release defects that are associated with each version in two steps. First, they mined the closed issues reports (i.e., post-release defects) that are associated with the studied projects in the JIRA Issue management system. Second, they linked these issue reports to their corresponding versions by using the earliest available version from the affected versions field that is provided in the issue report. Each closed issue report represents a post release defect. 

These chosen projects exhibit diversity along multiple dimensions. Such diversity helps us avoid conclusion instability and ensures that our obtained results generalize even on the datasets that are not a part of our study. We briefly explain below the various dimensions of diversity that our study exhibits.

\smallskip\noindent\textbf{Domain diversity.} We study projects that are from vastly different domains. For instance, we include projects that provide a NoSQL database (Hbase, Hive), search API (Lucene), an integration framework (camel) among our studied projects. The ``Description'' column in Table~\ref{table:dataset} provides a brief description of each of the studied projects. 

\smallskip\noindent\textbf{Size diversity.} We study projects that vastly differ in size. From Table~\ref{table:dataset}, we observe that the number of modules in our studied projects ranges from 559 to 8,714 and the total lines of code in our studied datasets range from 63.52k to 529.04K. 

\smallskip\noindent\textbf{Defective ratio diversity.} Our studied projects have defective ratios ranging from 2.20\% to 33.80\%. Such diversity ensures that we include projects of different quality in our study. 

\smallskip\noindent\textbf{Release diversity.} We study projects that include a variety of versions defined by different durations. The ``Duration (\#months)'' column in our Table~\ref{table:dataset} provides the number of elapsed months between the studied successive versions. From Table~\ref{table:dataset}, we observe that we have versions that have both shorter and longer elapsed time between the successive versions. For example in the ActiveMQ project, Version 5.1.0 was released in 2008-05-01, while version 5.2.0 was released almost after a year in 2009-06-30, and it was followed up with the next version 5.3.0 in another 4 months in 2009-10-23. While these are subsequent releases, we also include the Version 5.8.0 which was released 3 years in 2013-02-06 after Version 5.3.0 and has several other versions in between 5.3.0 and 5.8.0. Contrastingly, in the case of the HBase project, the duration between the studied versions are consistent and are only two months apart.

Finally, in addition to the code metrics that are provided by~\citet{yatish2019icse}, we extracted the SNA metrics for each version of each software project. To do so, we collected the source code of these projects. However, we weren't able to locate the source code for the versions Derby 10.3.1.4 and JRuby 1.7. Therefore, we do not include these two versions in our study.

\begin{table*}[!htp]
\centering
\caption{Overview of the studied software projects}\label{table:dataset}
\centering
\begin{threeparttable}
 
\begin{tabular}{p{30pt}p{102pt}p{40pt}rrrrrr} 
\hline
\textbf{Project} & \textbf{Description}&\textbf{Versions} & \textbf{KLOC} & \textbf{\#modules} & \textbf{\#edges} & \textbf{\#defective module} & \textbf{\#defects}&  \textbf{Duration (\#month)}\\
\hline
\multirow{5}{*}{\textbf{ActiveMQ}} &\multirow{5}{*}{Open source messaging server.} & 5.0.0 &141.42 & 1,869 &9,192& 291 (15.57\%)&603&2007-12-07\\
&&5.1.0&152.70&1,943&9,640&153 (7.87\%)&287&5\\
&& 5.2.0&167.50&2,011&9,961&218 (10.84\%)&478&14\\
&&5.3.0&159.74&2,011&11,968&231 (11.49\%)&681&4\\
&&5.8.0&228.26&2,339&18,957&173 (7.40\%)&346&39\\
\hline
\multirow{4}{*}{\textbf{Camel}} & \multirow{4}{*}{\tabincell{l}{Enterprise integration framework\\ to integrates systems producing\\ and consuming data.}} &1.4.0&74.21&1,481&7,608&280 (18.90\%)&430&2009-01-19\\
&&2.9.0&381.86&7,020&45,871&199 (2.83\%)&518&35\\
&&2.10.0&425.44&7,805&50,861&230 (2.95\%)&581&6\\
&&2.11.0&480.81&8,714&56,523&192 (2.20\%)&464&10\\
\hline
\multirow{2}{*}{\textbf{Derby}} &\multirow{2}{*}{\tabincell{l}{Relational database software.}} &10.2.1.6&410.24&1,941&20,347&656 (33.80\%)&2013&2006-10-06\\
&&10.5.1.1&529.04&2,672&27,318&383 (14.33\%)&1445&31\\
\hline
\multirow{3}{*}{\textbf{Groovy}}&\multirow{3}{*}{\tabincell{l}{A programming language}}&1.6.0.Beta1&71.38&608&3,275&63 (10.36\%)&142&2008-05-02\\
& &1.5.7&63.52&559&2,787&25 (4.47\%)&111&5\\
&&1.6.0.Beta2&78.73&670&3,794&75 (11.19\%)&184&5\\
\hline
\multirow{3}{*}{\textbf{HBase}} &\multirow{3}{*}{\tabincell{l}{A non-relational distributed\\ datasbase.}} &0.94.0&245.92&1,041&8,175&216 (20.70\%)&792&2012-05-14\\
&&0.95.0&503.63&1,600&12,308&378 (23.63\%)&1115&11\\
&&0.95.2&501.28&1,790&14,193&482 (26.92\%)&1262&4\\
\hline
\multirow{3}{*}{\textbf{Hive}} &\multirow{3}{*}{\tabincell{l}{A Data warehouse software built\\ on top of Hadoop.}} &0.9.0&225.15&1,311&8,595&247 (18.80\%)&377&2012-04-30\\
&&0.10.0&282.84&1,444&9,772&174 (12.05\%)&351&8\\
&&0.12.0&463.55&2,493&16,894&208 (8.34\%)&339&9\\
\hline
\multirow{3}{*}{\textbf{JRuby}} &\multirow{3}{*}{\tabincell{l}{The Ruby Programming Langua-\\ge
on the JVM.}} &1.1&104.71&689&5,849&87 (12.63\%)&173&2008-04-05\\
&&1.4&148.61&915&8,975&180 (19.67\%)&376&19\\
&&1.5&167.73&1,063&10,094&79 (7.43\%)&180&6\\
\hline
\multirow{4}{*}{\textbf{Lucene}} &\multirow{4}{*}{\tabincell{l}{Text search engine library.}} &2.3.0&101.10&798&3,051&195 (24.40\%) &428&2008-01-23\\
&&2.9.0&171.84&1,363&9,518&272 (19.96\%)&540&20\\
&&3.0.0&169.24&1,331&9,459&155 (11.65\%)&347&2\\
&&3.1.0&239.54&1,816&14,346&93 (5.12\%)&176&16\\
\hline
\multirow{3}{*}{\textbf{Wicket}}&\multirow{3}{*}{\tabincell{l}{Component oriented web applic-\\ation framework.}}&1.3.0-beta1&104.53&1,665&8,822&101 (6.07\%)&169&2007-04-23\\
&&1.3.0.beta2&108.34&1,756&9,372&130 (7.40\%)&194&3\\
&&1.5.3&164.95&2,565&14,737&105 (4.10\%)&244&51\\
\hline
\end{tabular}
 \begin{tablenotes}
 \scriptsize
\item In each of the project, in the "Duration (months)" column, for the first studied version, we only provide its release date. 
\end{tablenotes}
\end{threeparttable}
\end{table*}

\subsection{Experiment Setup}\label{sec:expsetup}

\subsubsection{Step 1: Metrics extraction}
To understand the effectiveness of the SNA metrics, we first extract the code and SNA metrics that we describe in Section~\ref{sec:metrics} for each version of the studied projects. 

\smallskip\noindent\textbf{Code metrics.} We use the code metrics by~\citet{yatish2019icse}. The 54 extracted code metrics for each version are provided in Table~\ref{table:metric}.

\smallskip\noindent\textbf{SNA metrics.} We extract the SNA metrics from the dependency network of a software system. We apply the Understand tool~\footnote{https://scitools.com/} to the source code of each version to construct the dependency network for each version of each project. Then we compute the SNA metrics of each node by using Ucinet tool~\footnote{https://sites.google.com/site/ucinetsoftware/home}. In total, we extract 40 EN metrics and 24 GN metrics as shown in Table~\ref{table:metric} in Section~\ref{sec:SNA}.


\subsubsection{Step 2: Out-of-sample Bootstrap sampling}
After we extract the metrics for each version of each studied project, we resample them 100 times with out-of-sample bootstrap sampling. We do so to ensure the statistical robustness of our results~\cite{Tantithamthavorn2016icse,tantithamthavorn2017tse}.


The out-of-sample bootstrap procedure resamples the data points from the dataset with replacement to create a \textit{train} set. The data points that were not sampled as a part of the \textit{train} set are set aside as the \textit{test} set. We repeat this procedure 100 times to create 100 \textit{train} sets and 100 \textit{test} sets. The generated \textit{test} sets have the same distributional properties as the \textit{train} set. 

Even when the \textit{test} sets can be obtained without such a resampling procedure, using the booststrap generated \textit{train} sets has several advantages. It helps us account for the non-deterministic nature of the used machine learning models and helps us ensure the  generalizability of the obtained results. For instance consider a commonly used SDP model like Random forest which is inherently non-deterministic. Therefore constructing the model just once on a given dataset might overestimate or underestimate the performance of the model. In addition, a bootstrap procedure helps us to non-parametrically approximate the underlying data distribution. By training our model on bootstrap generated \textit{train} sets, we ensure that we do not just draw our insights based on the model trained on just the studied dataset. Rather we ensure that the generated insights are based on the multiple samples drawn from the underlying distribution that generated the original dataset~\cite{Efron2000jasa,steyerberg2016jce,harrell2015regress}. In addition, as Harrell~\cite{harrell2015regress} argues performance estimates computed by the bootstrap procedure better quantify the variance in the computed performance estimates. 

Therefore, we use the \textit{train} sets to train the models and the \textit{test} sets to compute the performance measures. Depending on the SDP context, the composition of the dataset (the way the software projects/versions are grouped together) and the method of conducting the 100 out-of-sample bootstrap varies as follows:

\smallskip\noindent\textbf{Within-project context.} The \textit{train} and \textit{test} sets are generated by conducting a 100 out-of-sample bootstrap on each individual version. Each version of a project is considered as a dataset in the Within-project context.

\smallskip\noindent\textbf{Cross-version context.} The \textit{train} set is generated by conducting a 100 out-of-sample bootstrap on the dataset that is created by combining all versions of project that are released prior to the current version of the project. As described in Section~\ref{sec:sdp_contexts}, the \textit{train} set is generated by resampling the dataset made by combining versions V\textsubscript{1},V\textsubscript{2},\dots,V\textsubscript{k-1} of project A, where V\textsubscript{n} is the version of Project A for which defects are to be predicted. The \textit{test} remains the same in this context for all the 100 times. Therefore, a model is trained 100 times on 100 different resampled \textit{train} sets, but is tested 100 times against the same data points from the version V\textsubscript{k} of Project A. 


\smallskip\noindent\textbf{Cross-project context.} The \textit{train} set is generated by conducting a 100 out-of-sample bootstrap on the latest available version from one project (For example Project A). The latest available version of one project constitutes as the dataset in the Cross-project context similar to the studies by~\citet{Foyzur2013icse} and~\citet{Wan2016ist}. The \textit{test} set is created by considering each version of the target project (for example, version V\textsubscript{n} of Project B) i.e., the project for which the defects are to be predicted. Similar to the Cross-version context, while there are 100 different resampled \textit{train} sets, the \textit{test} set remains the same.

\subsubsection{Step 3: Data Pre-processing} In the Within-project and Cross-version contexts, as recommended by several prior studies~\cite{Arar2017asc, song2020tse}, we normalize the \textit{train} and \textit{test} sets. By doing so we ensure that all the studied metric values vary between 0 and 1.

In the Cross-project context, we perform a log transformation to standardize the studied metrics across the \textit{train} and \textit{test} sets. We do so as we follow the approach outlined by~\citet{cruz2009emse}. Such a procedure accounts for the concept drift that is common in the Cross-project context. We provide more details on why we choose the approach outlined by~\citet{cruz2009emse} is provided in the following step.

\begin{table*}[!htb]
\centering
\caption{The rankscore results as computed on the median performance measures across the bootstrap iterations.}\label{table:rank}
\hspace{0.5cm}
\centering
\begin{threeparttable}
\begin{tabular} {l|lrrrrrr} 
\hline
\textbf{Context}&\textbf{Measures} &\multicolumn{1}{l}{\textbf{ Code}} &\multicolumn{1}{l}{\textbf{ SNA}} & \multicolumn{1}{l}{\textbf{EN}} & \multicolumn{1}{l}{\textbf{GN}} &\multicolumn{1}{l}{\textbf{SM}} &\multicolumn{1}{l}{\textbf{COM}}\\
\hline
\multirow{14}{*}{\textbf{Within-Project}}&\multicolumn{7}{c}{\textbf{Defect-Count model}}\\
\cline{2-8}
&Adj R\textsuperscript{2} & 0.60 & 0.20 (2/30) & 0.40 (0/30) & 0.80 (9/30) & 0.00 (1/30) &\textbf{1.00} (11/30)\\

&RMSE & 0.60 & 0.40 (3/30) & 0.00 (0/30)& 0.20 (1/30)& \textbf{1.00} (10/30) &0.80 (7/30)\\

&Spearman & 0.60 & 0.20 (3/30) & 0.00 (0/30)&0.20 (1/30)&\textbf{1.00} (8/30)&0.80 (6/30)\\

&MAE & 0.40&0.20 (7/30)&0.00 (4/30)&0.60 (6/30)&\textbf{1.00} (13/30)&0.80 (10/30)\\
\cline{2-8}
&\multicolumn{7}{c}{\textbf{Defect-classification model}}\\
\cline{2-8}
&AUC &0.60&0.40 (4/30)&0.00 (0/30)&0.20 (4/30)&\textbf{1.00} (19/30)&0.80 (17/30)\\

&MCC &0.40&0.60 (7/30)&0.00 (3/30)&0.20 (5/30)&\textbf{1.00} (16/30)&0.80 (14/30)\\

&Recall & 0.20&0.40 (9/30) &0.00 (7/30) &0.60 (9/30) &\textbf{1.00} (15/30)&0.80 (12/30)\\

&Brier score & 0.60&0.40 (5/30)&0.00 (1/30)&0.20 (3/30)&0.80 (11/30)&\textbf{1.00} (13/30)\\

&Precision & 0.60&0.40 (3/30)&0.00 (0/30)&0.20 (2/30)&0.80 (9/30)&\textbf{1.00} (11/30)\\
\cline{2-8}
&\multicolumn{7}{c}{\textbf{Effort-aware model}}\\
\cline{2-8}
&ER&0.00&\textbf{1.00} (21/30)&0.80 (18/30) &0.40 (15/30)&0.20 (10/30)&0.60 (16/30)\\

&CE&0.00&\textbf{1.00} (21/30)&0.80 (19/30)&0.60 (18/30)&0.20 (14/30)&0.40 (16/30)\\
\hline
\multirow{14}{*}{\textbf{Cross-version}}&\multicolumn{7}{c}{\textbf{Defect-Count model}}\\
\cline{2-8}
&Adj R\textsuperscript{2} & \textbf{0.50} & 0.00 (4/21) & \textbf{0.50} (7/21) & 0.00 (4/21)& 0.00 (2/21)&\textbf{0.50} (12/21)\\

&RMSE & \textbf{1.00} & 0.50 (5/21)&\textbf{1.00} (7/21)& 0.50 (2/21)& \textbf{1.00} (7/21)&\textbf{1.00} (7/21)\\

&Spearman & \textbf{0.50} & 0.00 (1/21)& 0.00 (2/21)&0.00 (4/21)&\textbf{0.50} (5/21)&\textbf{0.50} (7/21)\\

&MAE & \textbf{1.00}&0.33 (0/21)&0.67 (1/21)&0.33 (0/21)&0.67 (1/21)&0.67 (5/21)\\
\cline{2-8}
&\multicolumn{7}{c}{\textbf{Defect-classification model}}\\
\cline{2-8}
&AUC &\textbf{0.50}&0.00 (2/21)&0.00 (3/21)&0.00 (3/21) &\textbf{0.50} (9/21)&\textbf{0.50} (10/21)\\

&MCC &\textbf{0.50}&\textbf{0.50} (7/21)&0.00 (4/21)&\textbf{0.50} (5/21)&\textbf{0.50} (8/21)&\textbf{0.50} (8/21)\\

&Recall & \textbf{0.50}&0.00 (4/21)&\textbf{0.50} (3/21)&0.00 (1/21)&\textbf{0.50} (6/21)&\textbf{0.50} (4/21)\\

&Brier score & 0.50&0.50 (9/21)&0.50 (8/21)&0.50 (6/21)&\textbf{1.00} (15/21)&0.50 (15/21)\\

&Precision & 0.00&\textbf{0.50} (13/21)&0.00 (4/21)&\textbf{0.50} (17/21)&\textbf{0.50} (12/21)&0.00 (10/21)\\
\cline{2-8}
&\multicolumn{7}{c}{\textbf{Effort-aware model}}\\
\cline{2-8}
&ER&0.00 &0.00 (6/21)&\textbf{0.50} (10/21)&0.00 (7/21)&0.00 (7/21)&\textbf{0.50} (11/21)\\

&CE&0.00&0.00 (5/21)&\textbf{0.50} (8/21)&0.00 (6/21)&0.00 (4/21)&0.00 (6/21)\\
\hline
\multirow{14}{*}{\textbf{Cross-Project}}&\multicolumn{7}{c}{\textbf{Defect-Count model}}\\
\cline{2-8}
&Adj R\textsuperscript{2} & \textbf{0.50} & \textbf{0.50} (0/30)& 0.00 (0/30)& \textbf{0.50} (0/30)& \textbf{0.50} (0/30)& \textbf{0.50} (1/30)\\

&RMSE & \textbf{1.00} & 0.50 (0/30)& 0.50 (0/30)& 0.50 (0/30)& \textbf{1.00} (0/30)&0.50 (1/30)\\

&Spearman & \textbf{0.50} & 0.00 (4/30)& 0.00 (4/30)& \textbf{0.50} (3/30)& \textbf{0.50} (3/30)&0.00 (1/30)\\

&MAE & \textbf{1.00}&0.50 (3/30) &\textbf{1.00} (3/30)&\textbf{1.00} (1/30)&0.50 (0/30)&\textbf{1.00} (2/30)\\
\cline{2-8}
&\multicolumn{7}{c}{\textbf{Defect-classification model}}\\
\cline{2-8}
&AUC &\textbf{0.50}&0.00 (0/30)&0.00 (1/30)&\textbf{0.50} (10/30) &0.00 (0/30)&\textbf{0.50} (2/30)\\

&MCC &\textbf{0.50}&0.00 (0/30)&\textbf{0.50} (4/30)&0.00 (0/30)&\textbf{0.50} (4/30)&\textbf{0.50} (4/30)\\

&Recall & \textbf{0.67}&0.33 (0/30)&0.33 (1/30)&0.33 (0/30)&\textbf{0.67} (3/30)&\textbf{0.67} (2/30)\\

&Brier score & 0.33&0.67 (12/30)&0.67 (6/30)&\textbf{1.00} (24/30)&0.33 (0/30)&0.67 (5/30)\\

&Precision & 0.00&0.00 (10/30)&\textbf{0.50} (13/30)&\textbf{0.50} (19/30)&0.00 (0/30)&0.00 (4/30)\\
\cline{2-8}
&\multicolumn{7}{c}{\textbf{Effort-aware model}}\\
\cline{2-8}
&ER&0.00&\textbf{0.50} (12/30)&\textbf{0.50} (14/30) &\textbf{0.50} (16/30)&0.00 (1/30)&0.00 (7/30)\\

&CE&0.00&0.00 (2/30)&0.00 (1/30)&\textbf{0.50} (10/30)&0.00 (0/30)&0.00 (2/30)\\
\hline
\end{tabular}
 \begin{tablenotes}
 \scriptsize
 \item Each value in the table is represented as R (x/y). 
 \item R- rankscore of the model trained on the metrics (given by the column) 
 \item x- No. of datasets on which the model trained on the metric family (given by the column) compared to the model trained on the code metrics in the given context have a larger than medium effect size. Please note that it is common for two metric families with similar rankscore R to have slightly different number of datasets)

 \item y- Total number of datasets for the given context in our study. The total number of datasets vary as the way in which the \textit{train} set is created varies based on context as outlined in Section~\ref{sec:sdp_contexts}.
 
 \end{tablenotes}
 \end{threeparttable}
\end{table*}



\subsubsection{Step 4: Model construction} \label{sec:construct_defect_model}
Models are constructed 100 times on the resampled datasets from each of the studied projects with hyperparameter tuning to ensure that our models fit the datasets well~\cite{tantithamthavorn2019tse}. Therefore, six models are constructed for each scenario on the six metric families (for details about metrics refer to Section~\ref{sec:metrics}) on each \textit{train} set for each project per bootstrap iteration. Therefore, for each SDP scenario, for each project, we build 600 models. 

We investigate three different SDP contexts as outlined in Section~\ref{sec:sdp_contexts} in our study. In order to ensure the validity of our findings, we choose the best baseline model in each of the studied SDP context to study the effectiveness of SNA metrics across the SDP contexts as follows:

\smallskip\noindent\textbf{Within-Project Context.} In the Within-project context, we use the Random Forest learner. We do so as~\citet{Baljinder2015icse} through a comprehensive benchmarking study, found Random Forest models to be among one of the best performing models in the Within-project context. In addition, Random Forest model is widely used by prior studies for the Within-Project defect prediction context~\cite{rajbahadur2017impact,yatish2019icse}. We use the~\texttt{RandomForestRegressor} function from the \texttt{Scikit-learn} Python package to build our Defect-count models, and the~\texttt{RandomForestClassifier} function from the \texttt{Scikit-learn} python package to build our Defect-classification and Effort-aware models. 
%
\smallskip\noindent\textbf{Cross-version Context.}  In the Cross-version context, we use the same Random Forest learner as the Within-project context. We do so as~\citet{amasaki2020ese} recently found that the difference in performance between the model constructed with the Random Forest learner and the top performing model in the Cross-version context (LACE2+NaiveBayes proposed by~\citet{peters2015icse}) to be small. Moreover, the Random Forest model is more accessible to practitioner and simpler than the LACE2+NaiveBayes model.


\smallskip\noindent\textbf{Cross-project Context.} In the Cross-project context, we use the log+NavieBayes model proposed by~\citet{cruz2009emse}. We do so as~\citet{steffen2019tse} found it to be the best performing model in the Cross-project context. The log+NaiveBayes model first performs a log transformation to standardize the studied metrics across the training and testing projects as we outline in the previous step. Following which, we build a Naive Bayes model on the log transformed data. We use the~\texttt{BayesianRidge} function from the \texttt{Scikit-learn} Python package to build our Defect-count models, and the~\texttt{GaussianNB} function from the \texttt{Scikit-learn} python package to build our Defect-classification and Effort-aware models. 



\smallskip\noindent\textbf{Hyperparameter tuning the studied models.} To ensure that the constructed models fit the datasets well, we use ~\texttt{GridSearchCV} from the \texttt{Scikit-learn\footnote{\url{https://scikit-learn.org/stable/modules/generated/sklearn.model_selection.GridSearchCV.html}}} Python package to tune the hyperparameters of learners. For Random Forest learner, we tune the ``n\_estimators" in the range $\{50, 100, 200, 500, 600\}$, and ``max\_depth" in the range $\{3, 5, 10, 15\}$. Similarly, for the Bayesian Ridge learner, we tune the the ``alpha\_1", ``lambda\_1", ``alpha\_2" and ``lambda\_2" in the range $\{1e-7, 1e-6, 1e-5, 1e-4, 1e-3\}$. The~\texttt{GaussianNB} function does not present any hyperparameters to tune. We tune these hyperparmaters for each bootstrap iteration. Then on each bootstrap iteration we choose the best performing hyperparameters for the learners.

\subsubsection{Step 5: Performance computation}
We compute the performance measures for the constructed models across the three studied SDP scenarios in each of the SDP contexts. We use the performance measures detailed in Section \ref{sec:measures}. For each SDP scenario, the performance of the model constructed on each of the six metric families is noted. In both the Within-project and Cross-version contexts, on each bootstrap iteration the 600 constructed models are tested on \textit{test} sets (100 models per \textit{train} and \textit{test} set for each studied metric family). These tests result in 600 performance scores for each of the studied performance measure. However, in the Cross-project context, the 600 models that are constructed for a given project are tested on each version present in each of the other 8 studied projects (since the \textit{test} set in a Cross-project context is each version from all the other studied projects). Therefore, for each project in the Cross-project context, 4,800 performance scores are computed (six metric families $\times$ 100 iterations $\times$ 8 projects).

\subsubsection{Step 6: Model evaluation}~\label{sec:model_evaluation}
To determine which metrics yield the best performing model for the studied SDP scenarios, we use a Friedman test with a post-hoc Nemenyi test, as suggested by~\citet{janez2006jmlr}. The Friedman test~\cite{friedman1940} checks if there are statistically significant differences between performance scores generated by the models trained on the studied metric families. If the Friedman test finds statistically significant differences, we employ the post-hoc Nemenyi test~\cite{nemenyi1963thesis} to obtain a ranking of the computed performance scores. Such a ranking helps us understand the metric family that generates the best performing models. 

However, the post-hoc Nemenyi test~\cite{nemenyi1963thesis} typically yields overlapping ranks~\cite{steffen2018tse}, which makes it harder to rank the effectiveness of the studied metrics. Therefore, we post process the results of the post-hoc Nemenyi test as suggested by~\citet{steffen2019tse} to generate a distinct \textit{rankscore} for each model trained on the studied metric families. The rankscore ranges between 0 and 1. A high rankscore for a model trained on a given metric family signifies that the given metric family generates the best performing model compared to the other metric families on the studied performance measure.


As mentioned in the previous step, for each dataset, 100 performance scores are computed for each constructed model on each of the metric families (In the Cross-project context, 800 performance scores, across the 8 other studied projects). Then for each dataset, we first compute the median of the performance scores. We then apply the Friedman test on the median performance scores across the studied datasets. If the results are significant, we then apply the post-hoc Nemenyi test to the generated performance scores. We then postprocess the results of the post-hoc Nemenyi test and generate the rankscore of each metric family for a given scenario across the studied datasets. We do so for all the studied SDP contexts and scenarios.

We also compute the Wilcoxon-signed rank test~\cite{qin2011tse} and Cliff's Delta effect size test~\cite{yang2017ist} between the models trained on code metrics and each of other studied metric families. However unlike the Friedman and post-hoc Nemenyi test, we compute the Wilcoxon-signed rank test and Cliff's delta effect size test per dataset (not across the studied datasets). Therefore we conduct the Wilcoxon-signed rank test and the Cliff's Delta effect size test directly on the bootstrap results of each of the studied datasets. That is across 100 bootstrap iterations for the Within-project and Cross-version context and 800 iterations for the Cross-project context. Moreover, since we perform multiple pairwise-comparisons, we correct the obtained p-values from the Wilcoxon-signed rank test with Bonferroni correction~\cite{richard2014opt} to control for false positives. We do so to statistically quantify the number of datasets on which models trained on other metric families outperform models trained on code metrics with a statistical significance (P-value$<$0.05) and a greater than medium effect size (i.e., if magnitude of Cliff's Delta $\geq0.33$).

\section{Results}\label{sec:case_study_results}
In this section, we present and discuss our study results related to our research questions. For all the RQs, we follow the experiment setup (as applicable to each SDP context that the RQ deals with) outlined in Section~\ref{sec:expsetup}.
\subsection{\textbf{(RQ1) \RQOne} }

\subsubsection{Results}
\smallskip\noindent\textbf{Result 1)} \textbf{\textit{In Within-project context, for both Defect-count and Defect-classification scenarios, models trained on SM and COM metrics outperform the models trained on other metric families.}}  Table~\ref{table:rank} lists the results of the rankscores (refer to Section~\ref{sec:model_evaluation}) for the models trained for all the SDP scenarios across the studied datasets. From Table~\ref{table:rank}, we observe that models trained on SM and COM metrics (both of which contain SNA metrics as a part of it) are consistently ranked at the top for Defect-count scenario. However, the number of datasets on which the models Defect-count models trained on SM and COM metrics outperform the models trained on code metrics with larger than medium effect size remain small (at most 13 out of the 30 studied datasets).

In contrast, for the Defect-classification scenario, we observe that on at least 9 of the 30 studied datasets, the Defect-classification models trained on SM and COM metrics outperform the models trained on code metrics with a larger than medium effect size. In particular, in terms of AUC, MCC and Recall, there are at least 15 out of the 30 studied datasets on which Defect-classification models trained on SM metrics outperform models trained on code metrics with statistical significance and a larger than medium effect size. Such results highlight the effectiveness of SNA metrics alongside with code metrics for the Defect-classification scenario.  


\smallskip\noindent\textbf{\textit{Result 2) Models trained on SNA metrics outperform models trained on other metric families for the Effort-aware scenarios across all the studied performance measures. Furthermore, at least on 21 of the 30 studied datasets Effort-aware models trained on SNA metrics outperform the models trained on code metrics with statistical significance and a larger than medium effect size.}} From Table~\ref{table:rank}, we also observe that models trained on EN and GN metrics are also consistently ranked above models trained on code metrics (with a statistical significance and a larger than medium effect size on at least 15 out of the 30 studied datasets). Such results highlight the effectiveness of SNA metrics for the Effort-aware scenario.  

\subsubsection{Analysis}\label{sec:informationanalysis}

From  the  above  results,  we  observe  that  models  trained  on  SM and COM metrics consistently outperform models trained on other metric families for Defect-count and Defect-classification scenarios. We hypothesize that such a result might be because SM and COM  metrics might capture richer information about software projects than the other metric families. We arrive at such a hypothesis as prior studies~\cite{Thomas2008icse,Thanh2010icsm,Wan2016ist,tosun2009promise} theorize that SNA metrics might provide information that is different or complimentary to code metrics. In addition, combing SNA metrics with code metrics could improve the performance of SDP models. 

In order to verify our hypothesis and quantify the information captured by the different metric families, we employ a Principal Component Analysis (PCA)-based analysis similar to~\citet{Baljinder2017msr}. We quantify the informativeness of each metric family in a given dataset by calculating the number of components that are needed to account for 95\% of the data variance information in each metric family. The higher the number of components that are required to account for 95\% of data variance information, the higher the informativeness of the metric family. We present the median number of components required by each studied metric family to account of 95\% of the data variance information across the  30  studied  datasets (presented  in  Table~\ref{table:dataset}) in Table~\ref{table:numberofpcs}.


\begin{table}[!htp]
\centering
\caption{The median number of components required by each metric family to account for 95\% of data variance information. 
}\label{table:numberofpcs}
\begin{tabular}{lrrrrrr}   
\hline
\textbf{Metric family} & \textbf{Code} & \textbf{SNA} & \textbf{EN} & \textbf{GN} & \textbf{SM} & \textbf{COM}\\
\hline
Median number&22&21&17&8&38&34\\
\hline
\end{tabular}
\end{table}

\smallskip\noindent\textbf{\textit{Result 3) The SM and COM metrics exhibit higher informativeness than the other metric families (see Table~\ref{table:numberofpcs})}}. Such a  result supports our  hypothesis that SNA metrics contained in the SM and COM  metrics adds additional information to the code metrics. Which in turn results in models trained on SM and COM metrics outperforming models trained on the other studied metric families. 

\subsection{\textbf{(RQ2) \RQTwo} }~\label{subsec:RQ2}



\subsubsection{Results}

\smallskip\noindent\textbf{\textit{Result 4)}} \textbf{\textit{In Cross-version context, Defect-count models trained on code metrics outperform the Defect-count models trained on other metric families across three out the four studied performance measures.}} From Table~\ref{table:rank} we observe that, the number of datasets on which Defect-count models trained on other metric families significantly outperform Defect-count models trained on code metrics in terms of Adj R\textsuperscript{2}, MAE, RMSE and Spearman is only at most 12. We note that Defect-count models trained on COM metrics (which combine SNA and code metrics) perform slightly better than the Defect-count models trained on code metrics according to Adj R$^2$ measure.

\smallskip\noindent\textbf{\textit{Result 5)}} \textbf{\textit{Defect-classification models trained on SM and COM metrics have a similar rankscore to that of models trained on code metrics.}}. On 3 of the studied performance measures (i.e., AUC, MCC, Recall) the Defect-classification models trained on SM and COM metrics have a similar rankscore to that of models trained on code metrics. However, Defect-classification models trained on SM metrics outperform Defect-Classification models trained on code metrics across 2 out of 5 studied performance measures (i.e., Brier score and Precision). On these two performance measures, from Table~\ref{table:rank}, we observe that Defect-classification models trained on SM metrics outperform the models trained on code metrics on at least 12 out of the 21 studied datasets with statistical significance and a larger than medium effect size. These results, demonstrate the potential usefulness of considering the SNA metrics alongside code metrics for the Defect-classification scenario in Cross-version context.

\smallskip\noindent\textbf{\textit{Result 6)}} \textbf{\textit{Effort-aware models trained on EN and COM metrics outperform Effort-aware models trained on other metric families.}} From Table~\ref{table:rank}, we observe that EN and COM metrics rank at the top. Interestingly, we observe that Effort-aware models trained on all the other metric families are ranked similarly. Such a result indicates that except for EN and COM metrics, other metric families yield Effort-aware models of similar performances. Such a result indicates that EN and COM metrics improve the performance of Effort-aware models.

\begin{table}[!htp]
\centering
\caption{The number of unchanged-status, changed-status and new modules per version across all the studied projects}\label{table:change}
\begin{threeparttable}
\begin{tabular}{llrrr}   
\hline
\textbf{Project} & \multicolumn{1}{l}{\textbf{Versions}} & \multicolumn{1}{l}{\textbf{\# Unchanged}} &\multicolumn{1}{l}{ \textbf{\#Changed}} & \multicolumn{1}{l}{\textbf{\#New}} \\
\hline
\multirow{4}{*}{\textbf{ActiveMQ}} & 5.1.0 &81.7\%&13.7\%&4.6\%\\
&5.2.0&84.3\%&10.0\%&5.7\%\\
& 5.3.0&72.2\%&10.7\%&17.1\%\\
&5.8.0&2.6\%&0.4\%&97.0\%\\
\hline
\multirow{3}{*}{\textbf{Camel}} 
&2.9.0&12.5\%&3.3\%&84.2\%\\
&2.10.0&86.5\%&2.8\%&10.7\%\\
&2.11.0&85.4\%&3.5\%&11.1\%\\
\hline
\multirow{1}{*}{\textbf{Derby}} &10.5.1.1&50.3\%&17.3\%&32.4\%\\
\hline
\multirow{2}{*}{\textbf{Groovy}} &1.5.7&83.6\%&6.1\%&10.3\%\\
&1.6.0.Beta2&83.7\%&5.4\%&10.9\%\\
\hline
\multirow{2}{*}{\textbf{HBase}} 
&0.95.0&40.9\%&13.6\%&45.5\%\\
&0.95.2&60.5\%&22.1\%&17.4\%\\
\hline
\multirow{2}{*}{\textbf{Hive}} 
&0.10.0&74.9\%&15.6\%&9.5\%\\
&0.12.0&50.0\%&6.3\%&43.7\%\\
\hline
\multirow{2}{*}{\textbf{JRuby}} 
&1.4&42.7\%&10.9\%&46.4\%\\
&1.5&70.0\%&14.9\%&15.1\%\\
\hline
\multirow{3}{*}{\textbf{Lucene}} 
&2.9.0&38.1\%&10.3\%&51.6\%\\
&3.0.0&87.6\%&10.8\%&1.6\%\\
&3.1.0&55.2\%&9.7\%&35.1\%\\
\hline
\multirow{2}{*}{\textbf{Wicket}} 
&1.3.0.Beta2&82.2\%&7.8\%&10.0\%\\
&1.5.3&18.1\%&1.3\%&80.6\%\\
\hline
\end{tabular}
 \begin{tablenotes}
   \scriptsize
  
   \item Unchanged - Unchanged-status modules.
   \item Changed - Changed-status modules.
   \item The first version from each software project is not included as it is not possible to identify the Unchanged-status,  Changed-status and new modules for the first version of the software where all the modules are New. 

 \end{tablenotes}
\end{threeparttable}
\end{table}

\begin{table*}[!htp]
\centering
\caption{The rankscore results as computed on the Defect-count model for Unchanged-status, Changed-status and New modules across all the studied datasets.}\label{table:rank_change}
\hspace{0.5cm}
\centering
  \begin{threeparttable}
   \begin{tabular} {l|rrrrrrrr}   

\hline
\multicolumn{2}{l}{\textbf{Measures}} &\multicolumn{1}{l}{\textbf{ Code}} & \multicolumn{1}{l}{\textbf{SNA}} & \multicolumn{1}{l}{\textbf{EN}} & \multicolumn{1}{l}{\textbf{GN}} & \multicolumn{1}{l}{\textbf{SM}} &\multicolumn{1}{l}{\textbf{COM}}\\
\hline
\multirow{4}{*}{\textbf{Unchanged-status}} &\multicolumn{1}{l}{Adj R\textsuperscript{2}} & \textbf{0.50}&0.00 (1/21)&0.00 (2/21)&0.00 (2/21)&0.00 (1/21)&\textbf{0.50} (7/21)\\

&\multicolumn{1}{l}{RMSE} &\textbf{1.00}&0.50 (1/21)&0.50 (1/21)&0.50 (1/21)&\textbf{1.00} (2/21)&\textbf{1.00} (4/21)\\

&\multicolumn{1}{l}{Spearman} &\textbf{0.50}&0.00 (1/21)&0.00 (1/21)&0.00 (1/21)&\textbf{0.50} (2/21)&\textbf{0.50} (5/21)\\

&\multicolumn{1}{l}{MAE} &\textbf{1.00}&0.33 (0/21)&0.67 (0/21)&0.33 (0/21)&0.67 (0/21)&0.67 (3/21)\\
\hline
\multirow{4}{*}{\textbf{Changed-status}} &\multicolumn{1}{l}{Adj R\textsuperscript{2}} & 0.33&0.33 (11/21)&\textbf{0.67} (18/21)&\textbf{0.67} (21/21)&0.00 (1/21)&\textbf{0.67} (21/21) \\

&\multicolumn{1}{l}{RMSE} &0.50&\textbf{1.00} (19/21)&\textbf{1.00} (17/21)&\textbf{1.00} (20/21)&\textbf{1.00} (19/21)&\textbf{1.00} (14/21)\\

&\multicolumn{1}{l}{Spearman} &0.00&\textbf{0.50} (16/21)&\textbf{0.50} (18/21)&\textbf{0.50} (18/21)&0.00 (6/21)&0.00 (8/21)\\

&\multicolumn{1}{l}{MAE} &0.50&\textbf{1.00} (18/21)&\textbf{1.00} (20/21)&\textbf{1.00} (18/21)&0.50 (8/21)&0.50 (6/21)\\
\hline
\multirow{4}{*}{\textbf{New}} &\multicolumn{1}{l}{Adj R\textsuperscript{2}} & \textbf{0.50}&0.00 (3/21)&\textbf{0.50} (12/21)&\textbf{0.50} (13/21)&0.00 (4/21)&\textbf{0.50} (15/21)\\

&\multicolumn{1}{l}{RMSE} &\textbf{1.00}&\textbf{1.00} (8/21)&\textbf{1.00} (7/21)&\textbf{1.00} (7/21)&\textbf{1.00} (10/21)&\textbf{1.00} (10/21)\\

&\multicolumn{1}{l}{Spearman} &\textbf{1.00}&\textbf{1.00} (5/21)&\textbf{1.00} (5/21)&\textbf{1.00} (8/21)&\textbf{1.00} (8/21)&\textbf{1.00} (8/21)\\

&\multicolumn{1}{l}{MAE} &\textbf{1.00}&0.50 (4/21)&\textbf{1.00} (8/21)&0.50 (4/21)&0.50 (6/21)&\textbf{1.00} (10/21)\\
\hline












\end{tabular}

\end{threeparttable}
\end{table*}

\subsubsection{Analysis}
From the above results, we observe that for the Defect-count scenario, code metrics are more effective than the other metric families. Such a result is in contrast with the observed results in the Within-project context and other SDP scenarios in the Cross-version context. For the other SDP scenarios of the Cross-version context, where SNA (particularly, EN metrics) metrics by themselves or in combination with code metrics improve the performance of models over models built using code metrics only. We hypothesize that such a result might be because, as software systems evolve, the defective status of a large number of modules across multiple versions remains unchanged due to a couple of reasons. First, the modules that had defect(s) in one version tend to be defective across multiple versions~\cite{jacek2005msr}. Therefore, some modules remain defective across multiple versions. Second, many modules are not modified across multiple versions, therefore their defective status typically remains unchanged (especially for non-defective modules). For instance, many of the modules which were not defective in version 1.0 are typically not modified for version 1.1 (since they are already performing their intended function without any defects) therefore their defective status remains ‘non-Defective’ across both versions 1.0 and 1.1. Therefore only a few modules change their defective status across versions. We refer to these modules whose defective status remains unchanged across versions as “unchanged-status modules”. Because of the aforementioned reasons, a large number of the data points have exactly the same defective status  across two successive versions (i.e., the train and test data). 

As~\citet{Foyzur2013icse} show that code metrics typically have a higher stasis (i.e., they do not change much) on unchanged-status modules. Which would imply that there exists a large number of modules across successive versions  (i.e., the train and test data)  which are more similar in terms of code metrics across successive releases than in terms of the SNA metrics. In turn, Defect-count models trained on code metrics could easily predict the defective status of these unchanged-status modules. In other words, the models trained on just code metrics would repeatedly predict the same modules to be defective over and over again across the versions. Therefore, if the number of unchanged-status modules across versions is high, then the performance of code metrics on these modules might dominate the performance of the Defect-count models trained on metric families which contain SNA metrics. Furthermore, if the Defect-count models trained on code metrics perform well only on the modules that exhibit the same defective status across versions, and not on the other modules, the SNA metrics might still be useful by themselves are in combination with code metrics in predicting the number of defects in Cross-version context.

To validate our hypothesis, we do three things: First we compute the number of  software modules that have the same defective status (unchanged-status  modules) across each version of a given project. We also compute the number of modules that changed their defective status (changed-status modules) and the number of newly added modules (new modules). We do so to identify if there exists a significant number of unchanged-status modules across each version for the studied software projects.

Second, we verify if the stasis of code metrics are indeed higher than that of SNA metrics for these unchanged-status modules. To do so, we compare the stasis of code and SNA metrics on the unchanged-status modules across each version of the studied projects. We compute the stasis of the metrics using the approach outlined by~\citet{Foyzur2013icse}. We compute the spearman correlation between each metric from both the code and SNA metric families across each two successive versions and average them. We present these average values as the stasis of code and SNA metric families respectively. We do so to see if the superior performance of Defect-count models trained on code metrics are due to the high stasis of code metrics on unchanged-status modules.


Third, for each Defect-count model trained on the Cross-version train set (refer Section 3.2.2), we split the Cross-version test set into unchanged-status, changed-status, and new software modules. We then calculate the performance of the Defect-count models on them separately. We do so for the Defect-count models trained on all the metric families across all the studied projects. We do so to identify if the Defect-count models trained on code metrics outperform other studied metric families only on the  unchanged-status modules or across all the types of modules. If the code metrics outperform the SNA metrics only on  unchanged-status modules (and does not outperform the other metric families on the changed-status modules) then such a result would support our hypothesis and vice versa.

\textbf{Except for ActiveMQ~5.8.0, Camel~2.9.0, HBase~0.95.0, JRuby~1.4, Lucene~2.9.0 and Wicket~1.5.3, all the versions across the studied projects (15 of 21) have 50-87.6\% unchanged-status modules.} Table~\ref{table:change} shows the number of unchanged-status, changed-status and new software modules per version. The results in Table~\ref{table:change} supports our hypothesis that the presence of a high number of  unchanged-status modules favours the Defect-count models that are trained on code metrics. 

\textbf{Across all the studied projects, we observe that the code metrics exhibit significantly higher stasis than SNA metrics for the  unchanged-status modules.} Such a result, along with the high number of unchanged-status modules across the different versions, explain the superior performance of Defect-count models trained on code metrics in the Cross-version context. We compute the stasis values of the code and SNA metrics in the  unchanged-status modules across the studied projects, and observe that code metrics exhibit an extremely high median stasis of 0.97 on the unchanged-status modules with a very limited range between 0.6 and 1. In contrast, the SNA metrics have a median stasis of 0.86 and show a significantly wider range varying from -1 to 1. Furthermore, a Wilcoxon Signed Rank test between the computed stasis values of the code metrics and SNA metrics of the  unchanged-status modules across the studied project exhibits a statistically significant difference (p-value $<$ 0.05) with a small cliff’s delta effect size of 0.3.
 

\smallskip\noindent\textbf{\textit{Results 7) Defect-count models trained on code metrics do not outperform models trained on other metric families for changed-status modules across all the studied versions and performance measures.}} Table~\ref{table:rank_change} shows the results of the rankscore for the Defect-count models trained on unchanged-status, changed-status and new modules for all the studied metric families. We note that, while the Defect-count model trained on code metrics performs well on unchanged-status and new modules, it does not perform as well on changed-status modules. These results in addition with the high number of  unchanged-status and new modules for each version across the studied projects could potentially indicate the reason for the superior overall performance of the Defect-count models trained on code metrics.

\smallskip\noindent\textbf{\textit{Results 8) Defect-count models trained on EN, GN and SNA metrics are useful at identifying the number of defects in changed-status modules across the software versions on three of the four studied performance measures.}} In addition, From Table~\ref{table:rank_change}, we observe that across all the performance measures on at least 11 out of the 21 studied datasets, Defect-count models trained on EN, GN and SNA metrics significantly outperform Defect-count models trained on other metric families (including code metrics). Such a result indicates that SNA metrics indeed improve the performance for the Defect-count scenario in the Cross-version context. Therefore these results reiterate the need for considering SNA metrics even as a part of the Defect-count scenario, where at first glance, it appears that SNA metrics might not be as effective.

\subsection{\textbf{(RQ3) \RQThree} }\label{subsec:RQ3}


\subsubsection{Results}

\smallskip\noindent\textbf{\textit{Result 9) In the Cross-project context, Defect-count models trained on the code metrics outperform the Defect-count models trained on other metric families across all the studied performance measures.}} From Table~\ref{table:rank} we observe that the number of datasets on which the Defect-count models trained on other metric families that outperform the Defect-count models trained on code metrics is less than 4. Even the SM and COM metrics, which contain code metrics are not as effective for the Defect-count scenario in the Cross-project context. 



\smallskip\noindent\textbf{\textit{Result 10) Defect-classification models trained on the studied SNA metrics do not outperform the Defect-classification models trained on code metrics on 3 out the 5 studied performance measures (i.e., AUC, MCC and Recall).}} Across these three performance measures, only Defect-classification models trained on COM metrics have similar performance (not higher) as the Defect-classification models trained on code metrics. 

Defect-classification models trained on GN metrics outperform the Defect-classification models trained on all the other metric families on  2 out of 5 the studied performance measures. From Table~\ref{table:rank} we observe that the models trained on GN metrics exhibit superior performance than models trained on code metrics with a larger than medium effect size on at least 19 out of the 30 studied datasets in terms of Brier-score and Precision. Such a result showcases that while there might be some slight advantage that can be potentially realized by using SNA metrics (i.e., GN metrics). However, we cannot conclude that using SNA metrics would consistently yield better results than using training models with just code metrics for Defect-classification scenario in the cross-project context.



\smallskip\noindent\textbf{\textit{Result 11) Effort-aware models trained on GN metrics outperform the Effort-aware models trained on all the other metric families.}} From Table~\ref{table:rank} we observe that the models trained on GN metrics exhibit superior performance than models trained on code metrics with a larger than medium effect size on atleast 10 out of the 30 studied datasets.

\subsubsection{Analysis}
Models trained on code metrics outperform models trained on other metrics for the Defect-count and Defect-classification scenario. Such a result is not particularly surprising as~\citet{hosseini2017systematic} through a meta analysis find that models trained on code metrics typically perform the best in the Cross-project context. However, in contrast to the findings of~\citet{hosseini2017systematic} we find that models trained on GN metrics outperform the models trained on code metrics for Defect-classification (on 2 out of the 5 studied performance measures) and Effort-aware scenarios. In addition, for the Defect-classification scenario, on the MCC and Recall measures, models trained on SM and COM metrics (which contain SNA metrics) perform at least as well as models trained on code metrics, if not outperform them.
 
We argue that the different information in the SNA metrics (in particular the GN metrics) help break the ceiling effect~\cite{menzies2008implications} exhibited by code metrics. As~\citet{menzies2008implications} explain, code metrics contain only limited information content. To improve the performance of models in the Cross-project context,~\citet{herbold2017comparative} argue that metrics other than code metrics need to be considered to train Defect-classification models. They argue that other metrics might help us break the ceiling effect exhibited by code metrics due to their limited information content. Since the SNA metrics capture information that is not typically captured by the code metrics, we argue that SNA metrics (in particular the GN metrics) help break the ceiling effect and in turn provides comparable and at times better performance for the Defect-classification and Effort-aware scenarios.

\section{\textbf{Discussion} }
Table~\ref{table:result} summarizes the results that we observed from Section~\ref{sec:case_study_results}. From Table~\ref{table:result}, we see that SNA metrics are useful in all the studied SDP contexts and several of the studied SDP scenarios. Such a result is in contrast with results presented in Table~\ref{table:survey}. Table~\ref{table:survey} highlights that there is no clear consensus on the effectiveness of SNA metrics over the code metrics across the different SDP contexts and scenarios. In this section, we highlight potential reasons why prior studies portray a mixed picture of the effectiveness of SNA metrics over code metrics across the three SDP contexts. 

\renewcommand{\arraystretch}{1.5}
\begin{table*}[tb]
\caption{The observed effectiveness of SNA metrics over code metrics across the studied SDP scenarios and contexts in our study.}\label{table:result}
\begin{tabular}{l|c|c|c|c|c|c|c|c|c} 
\hline
\multirow{2}{*}{\textbf{Metric family}}& \multicolumn{3}{c|}{\textbf{Within-project}} & \multicolumn{3}{c|}{\textbf{Cross-version}} &\multicolumn{3}{c}{\textbf{Cross-project}}\\
\cline{2-10}
& \multicolumn{1}{c|}{\textbf{D-classification}} & \multicolumn{1}{c|}{\textbf{D-count}}& \multicolumn{1}{c|}{\textbf{E-aware}} &\multicolumn{1}{c|}{\textbf{D-classification}} & \multicolumn{1}{c|}{\textbf{D-count}}& \multicolumn{1}{c|}{\textbf{E-aware}} &\multicolumn{1}{c|}{\textbf{D-classification}} & \multicolumn{1}{c|}{\textbf{D-count}}& \multicolumn{1}{c}{\textbf{E-aware}}\\
\hline
\textbf{Code metrics} &&&&$\surd$&$\surd$&&$\surd$&$\surd$&\\
\hline
\textbf{SNA metrics} &&&$\surd$&&&&&&\\
\hline
\textbf{EN metrics} &&&&&&$\surd$&&& \\
\hline
\textbf{GN metrics} &&&&&&&&&$\surd$ \\
\hline
\textbf{SM metrics} &$\surd$&$\surd$&&&&&&\\
\hline
\textbf{COM metrics} &$\surd$&$\surd$&&&&$\surd$&&&\\
\hline
\end{tabular}
\end{table*}

\subsection{Within-project context}
 
Prior studies that do not find SNA metrics to be more effective than code metrics in the Within-project context only use a small subset of SNA metrics in their studies. From Table~\ref{table:result} we clearly observe that the models trained on  SM and COM metric families perform better than models trained on code metrics across the Defect-count and Defect-classification scenario. For the Effort-aware scenario, SNA metrics prove to be more effective than code metrics.  While most prior studies come to a similar conclusion as ours~\cite{Premraj2011esem,Thomas2008icse,Wan2016ist,Thanh2010icsm},~\citet{Satya2013apsec} and~\citet{Fang2014js} find code metrics to be equally or more effective than SNA/SM metrics. We argue that these studies come to a different conclusion as they used only a small subset of the available SNA metrics in their studies.~\citet{Satya2013apsec} used only 12 SNA metrics and~\citet{Fang2014js} used only five SNA metrics in their studies (in contrast we used 64 SNA metrics). However the studies that found SNA metrics to be more effective than code metrics typically used more than 60 SNA metrics (including our study).

\subsection{Cross-version and Cross-project contexts}

We argue that prior studies do not observe the effectiveness of SNA metrics over code metrics in both the Cross-version and Cross-project contexts because they do not consider EN and GN metrics separately. As we observe from Table~\ref{table:survey}, prior studies~\cite{Premraj2011esem,Wan2016ist,Satya2013apsec} state that the models trained on code metrics perform similarly or better than the models trained on SNA metrics. They do so for the  Effort-aware scenarios of both the Cross-version and Cross-project contexts. From Table~\ref{table:rank} we could observe similar findings if we only consider SNA and SM metrics in our study. However, the effectiveness of the SNA metrics in the Cross-version and Cross-project contexts can be observed only when the GN, EN and COM metrics are considered separately (as we can see from Table~\ref{table:result}). 

We observe that considering these metric families separately showcase the effectiveness of SNA metrics because GN and EN metrics capture different characteristics of a software system. We argue that considering GN and EN metrics separately helps us avoid Simpson's paradox~\cite{Clifford1982stat}. Simpson’s paradox is a phenomenon where combining separate views of data does not necessarily result in a more informative grouping. Rather, it might in fact weaken individual effects exhibited by each view~\cite{Clifford1982stat}. In fact, such a grouping may even present a signal that is not faithful to either of the underlying views~\cite{Clifford1982stat}. In our case, GN metrics capture the dependency characteristics of a given module in the context of the whole software system (i.e., a global view). Whereas, EN metrics capture the dependency characteristics of a given software module with respect to its neighboring nodes (i.e., a local view). Therefore, we argue that considering the GN and EN metrics together as SNA metrics might weaken the strength of the signal that they may individually contain. Therefore, we argue that to avoid Simpson’s paradox, future studies should not only consider the combined effects of different views of data. Instead, they should also separately consider the impact of each view by itself in addition to the combined view. 

\textbf{We argue that prior studies do not clearly capture the effectiveness of SNA metrics because of two key reasons: 1) Some studies only consider a small subset of the available SNA metrics in their study 2) They do not consider the GN and EN metrics separately in their studies.}

\section{Implications}\label{sec:implications}
In this section, we outline the implications that can be inferred from our results. We present these implications to guide researchers and practitioners who seek to build SDP models in the future. 

\smallskip\noindent\textbf{\textit{Implication 1) Researchers and practitioners should consider using SNA metrics by themselves or alongside code metrics when building models for any of the studied SDP scenarios.}} From the results presented in Section~\ref{sec:case_study_results}, we observe that SNA metrics improve the performance of models for 5 out of the 9 studied SDP scenarios (three SDP scenarios across three SDP contexts). More specifically, except for the Defect-count and Defect-classification scenario for both Cross-version and Cross-project SDP contexts, we find that models trained on SNA metrics by themselves (GN or EN metrics) or SNA metrics alongside code metrics (SM or COM metrics) perform equally or at least slightly better (if not more).

\textbf{However, none of SDP studies in the recent years consider SNA metrics in their studies when building models.} For instance, consider these handful of studies published in ICSE or TSE 2018 and 2019. None of these studies that built models for the Within-project, consider the SNA (or GN metrics)~\cite{yatish2019icse, tantithamthavorn2019tse, Hosseini2019tse, Li2019tse, Sohn2019tse}. In addition, the effort of computing SNA metrics is not significantly greater than computing just the code metrics. For instance, prior studies~\cite{yatish2019icse} typically use the Understand tool to compute code metrics. To compute SNA metrics, as we explain in Section~\ref{sec:expsetup}, in addition to using the Understand tool, we use the UciNet tool. Therefore, given the potential performance improvements and the ease of computing the SNA metrics, we urge the researchers and practitioners to consider using SNA metrics alongside the code metrics in their SDP studies.

\smallskip\noindent\textbf{\textit{Implication 2) EN and GN metrics should be considered separately.}} EN and GN metrics have different impact on models depending on the context and scenario. For instance, in the Cross-project context, for the Effort-aware scenario,  models trained on just GN metrics have better performance compared to models trained on other metric families. Furthermore, in the Cross-version context, Effort-aware models trained on the EN metrics exhibit improved performance over the models trained on other metric families. These results indicate the necessity of considering the EN and GN metrics separately for SDP.
\section{Threats to validity}\label{sec:threats_to_validity}
In this section, we discuss the threats to validity of our presented results and inferences.

\smallskip\noindent\textbf{Construct Validity.} The number of observed post-release defects in our studied datasets and the version in which the defects were introduced might not be accurate. Da Costa et al.~\cite{da2017tse} suggested that the realistic estimation of the earliest affected version for a given defect depends heavily on development team of the software project. In practice, a development team may provide incomplete or wrong estimation of the number of defects or the version in which the defect was originally introduced. Such inconsistencies may generate false negatives. To mitigate such instances, we construct our experiments on a dataset provided by~\citet{yatish2019icse}, where they calculated the number of post-release defects for each module with a realistic approach over the traditionally used heuristic approach. The used approach seeks to eliminate some of the associated noise with aforementioned inconsistencies. However, some amount of noise in the dataset cannot be avoided.~\citet{Tantitham2015icse} observed that such noise impacted the recall of the SDP classifiers. Therefore, we suggest that future studies should revisit the effectiveness of the SNA metrics across various SDP contexts and scenarios with an approach similar to ours using datasets devoid of noise.

Another threat to construct validity is that we do not consider multiple methods outlined by~\citet{Feng2017tse} to aggregate the code metrics at the file-level. 
~\citet{Feng2017tse} showed when building Defect-classification models, using code metrics aggregated through several aggregation schemes is better than using summation. However, in our study, though we do not use all the aggregation methods that~\citet{Feng2017tse} used to aggregate code metrics, we still use the mean, min and max methods of aggregation in addition to the summation method. We observe that our use of several aggregation methods helps mitigate the threat outlined by~\citet{Feng2017tse} for the Defect-classification scenario. Furthermore, for the case of Defect-count and Effort-aware scenarios,~\citet{Feng2017tse} showed that using code metrics aggregated through summation is as effective as using code metrics aggregated through several different aggregation schemes. Therefore, we argue that our results are still relevant to the current state-of-practice. However, we acknowledge that investigating the effectiveness of SNA metrics over code metrics aggregated through several aggregation schemes for the Defect-classification scenario is an interesting avenue for future research.

Another  threat  to  construct  validity  is  that  we quantify the informativeness of each metric family by calculating the  number  of  components  that  are  needed  to  account  for  95\% of  the  data  variance  information similar to~\citet{Baljinder2017msr}. However, on the off chance that any studied metric family is purely composed of random values, our procedure for calculating the informativeness of the metric families could be impacted by that metric family composed of random values. As such, it presents a threat to the analysis that we conduct in Section~\ref{sec:informationanalysis}. However, such a result is unlikely, since the metric families that we find as the most informative (i.e., SM and COM metrics) are indeed the best performing metrics across all the studied datasets for the Within-project context. Such a result would not be possible if the SM and COM metrics were random values. Nevertheless, we encourage the future studies to revisit the informativeness of our studied metrics with more robust methods of calculating informativeness.

Finally, prior studies show that the results obtained from Defect-count and Defect-classification models could potentially be influenced by the module size~\cite{koru2005promise}. In particular, if the size of a module heavily correlates with the number of defects or defectiveness of a module, then a more trivial  SDP models could simply predict large modules as defective. Therefore, to investigate if such a threat exists in our studied projects, we compute the spearman correlation between the lines of code and the number of defects in a module, across all the studied datasets. We observe that the lines of code metric is not highly correlated with the number defects across the studied datasets. We observe that the median spearman correlation across all the studied datasets is only 0.3. Such a result asserts that the Defect-count and Defect-classification models do not rely solely on the size of the inspected modules in our studied projects.

\smallskip\noindent\textbf{External Validity.} We analyze the effectiveness of the SNA metrics over code metrics on nine fixed open source software projects developed in  JAVA.  Though the studied projects are diverse,  there  is  a  threat  that  our  findings  may  not  generalize for  projects with different module sizes and version definitions. In addition, as~\citet{Feng2016esme} even if different software projects appear seemingly different, their software metrics might be similarly or differently distributed depending on several contextual features. Therefore it is possible that the observed usefulness of SNA metrics on our studied projects might not generalize for projects that are of a different nature. Therefore, we encourage future studies to explore the impact of the nature, module size and version definitions of a studied software project on the usefulness of SNA metrics across different SDP contexts and scenarios.

Another threat to external validity is that we only use Area Under the Cost-Effectiveness Curve and Effort Reduction measures  (which  rely  only  on  LOC)  to  quantify  effort.~\citet{herbold2019tse} specified a cost model that takes more than LOC into account to quantify effort. We argue that our studied datasets provided by \citet{yatish2019icse} only include the static code metrics and number of defects present in each module, which make our study unable to adapt the cost model. However, we invite future research to explore the use of this cost model to understand the effectiveness of SNA metrics for the Effort-aware scenarios.

Finally, we recommend researchers and practitioners to consider SNA metrics alongside code metrics in their SDP studies. Considering both the metric families together, particularly when building SDP models for small projects with limited data points might result in over fitted SDP models. These SDP models might not perform well in practice and as such it is a threat to the external validity of our recommendation. To mitigate this threat, we urge the researchers and practitioners to use robust feature selection methods~\cite{chen2006survey} and model validation methods like out-of-sample bootstrap~\cite{tantithamthavorn2017tse} as a part of the experimental setup that they use to construct their SDP models.

\smallskip\noindent\textbf{Internal validity. } A key threat to internal validity is that we only study the effectiveness of SNA metrics compared to code metrics for SDP and do not consider other metric families like process metrics.~\citet{Foyzur2013icse} showed the models built on process metrics outperform the models built on code metrics. Similarly several other metric families show the superiority of change metrics~\cite{Marco2009re, graves2000tses, liu2017esem}, churn metrics~\cite{Nachiappan2005icse, Nachiappan2007esem} for SDP. However, we do not consider other metrics, as in the absence of history of a software project, only the source code of a software is available. In cases such as those, we can only extract code metrics and SNA metrics. Therefore, we argue that our findings help researchers and practitioners better leverage the source code for SDP. Especially, when the history of a project is not available. Furthermore, we also consider all combinations of the metrics extracted from the source and thereby eliminate potential confounders. However, we invite the future research to understand the effectiveness of SNA metrics in the context of the other commonly used metric families in SDP.

\section{Conclusions}\label{sec:conclusions}
The effectiveness of the SNA metrics for models across the common SDP contexts and scenarios has been widely debated. Some prior studies observe that SNA metrics are more effective than the widely used code metrics~\cite{Thomas2008icse,Thanh2010icsm}, whereas several other prior studies observe that SNA metrics are no more effective than the code metrics~\cite{Satya2013apsec}. Therefore it is pivotal to understand the effectiveness of SNA metrics in comparison to code metrics across the various common SDP contexts and scenarios. As these are the only few metrics other than code metrics that can be extracted when the history of a software project is not available (e.g., new project or not historical tracking). 

Therefore, in this paper, we revisit the effectiveness of SNA metrics for the common SDP scenarios in different SDP contexts through a case study on 30 versions of 9 open source software projects. The results of our case study suggest that considering SNA metrics by themselves or along with the code metrics could potentially improve the performance of models for at least 5 out of the 9 studied SDP scenarios (three SDP scenarios across three SDP contexts). However, we do note that in some cases, the improvements in performance by using SNA metrics alongside or over simply using code metrics might be marginal. Whereas in other cases it might be large.  



We do not claim the generalization of our observed results. Instead, we simply observe that there exist software projects in which using SNA metrics by themselves or along with code metrics could yield better performing SDP models across several SDP contexts and scenarios. Hence, we suggest that future research should consider using SNA metrics alongside code metrics when building SDP models for various SDP contexts and scenarios. Finally, we suggest that EN and GN metrics (that make up the SNA metrics) should be considered separately along with being considered together.

\ifCLASSOPTIONcompsoc
  \section*{Acknowledgments}
\else
  \section*{Acknowledgment}
\fi

This paper would not have been possible without the generous support by the Fundamental Research Funds for the Central Universities under grant No.2019XKQYMS84.

\ifCLASSOPTIONcaptionsoff
  \newpage
\fi

\bibliographystyle{IEEEtranSN}
\begin{footnotesize}
\balance
\bibliography{bibliography}
\end{footnotesize}
\vspace{-1cm}
\begin{IEEEbiography}[{\includegraphics[width=1in,height=1.25in,clip,keepaspectratio]{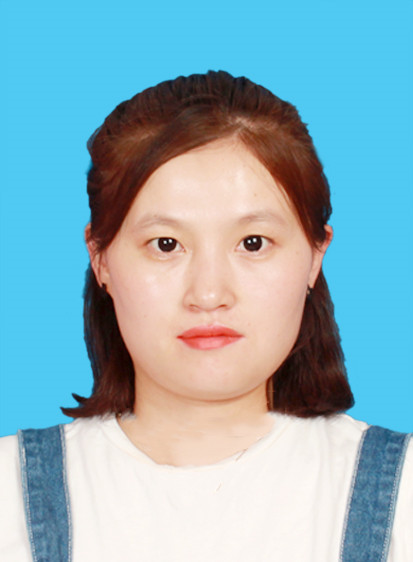}}]{Lina Gong}
is currently a lecturer in the School of Computer Science and Technology, Nanjing University of Aeronautics and Astronautics, China.
She received her Ph.D. degree in the Computer software and theory from China University of Mining and Technology, China, her BE degree
in Software Engineering from China University of Petroleum, China. She also studied as a visitor one year in the Software Analysis and
Intelligence Lab (SAIL), School of Computing, Queen’s University, Canada. Her research interests include machine learning, software analysis, software testing and mining software repositories.

\end{IEEEbiography}
\vspace{-1cm}
\begin{IEEEbiography}[{\includegraphics[width=1in,height=1.25in,clip,keepaspectratio]{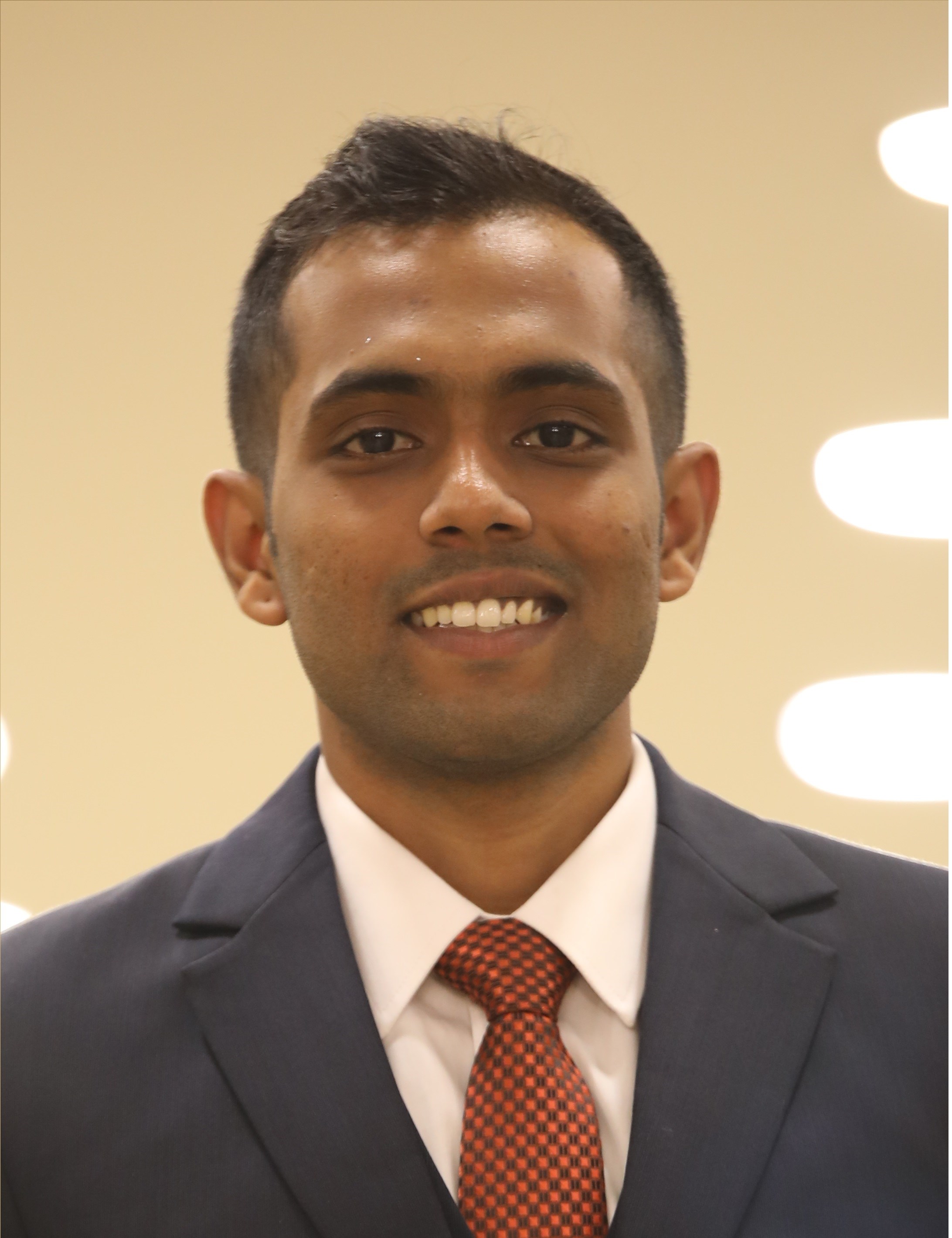}}]{Gopi Krishnan Rajbahadur} holds a PhD in computer science from Queen's University, Canada. He received his BE in computer Science and Engineering from SKR Engineering college, Anna University, India. His research interests include Software Engineering for Artificial Intelligence (SE4AI), Artificial Intelligence for Software Engineering (AI4SE), Mining Software Repositories (MSR) and Explainable AI (XAI).
\end{IEEEbiography}
\vspace{-1.1cm}
\begin{IEEEbiography}[{\includegraphics[width=1in,height=1.25in,clip,keepaspectratio]{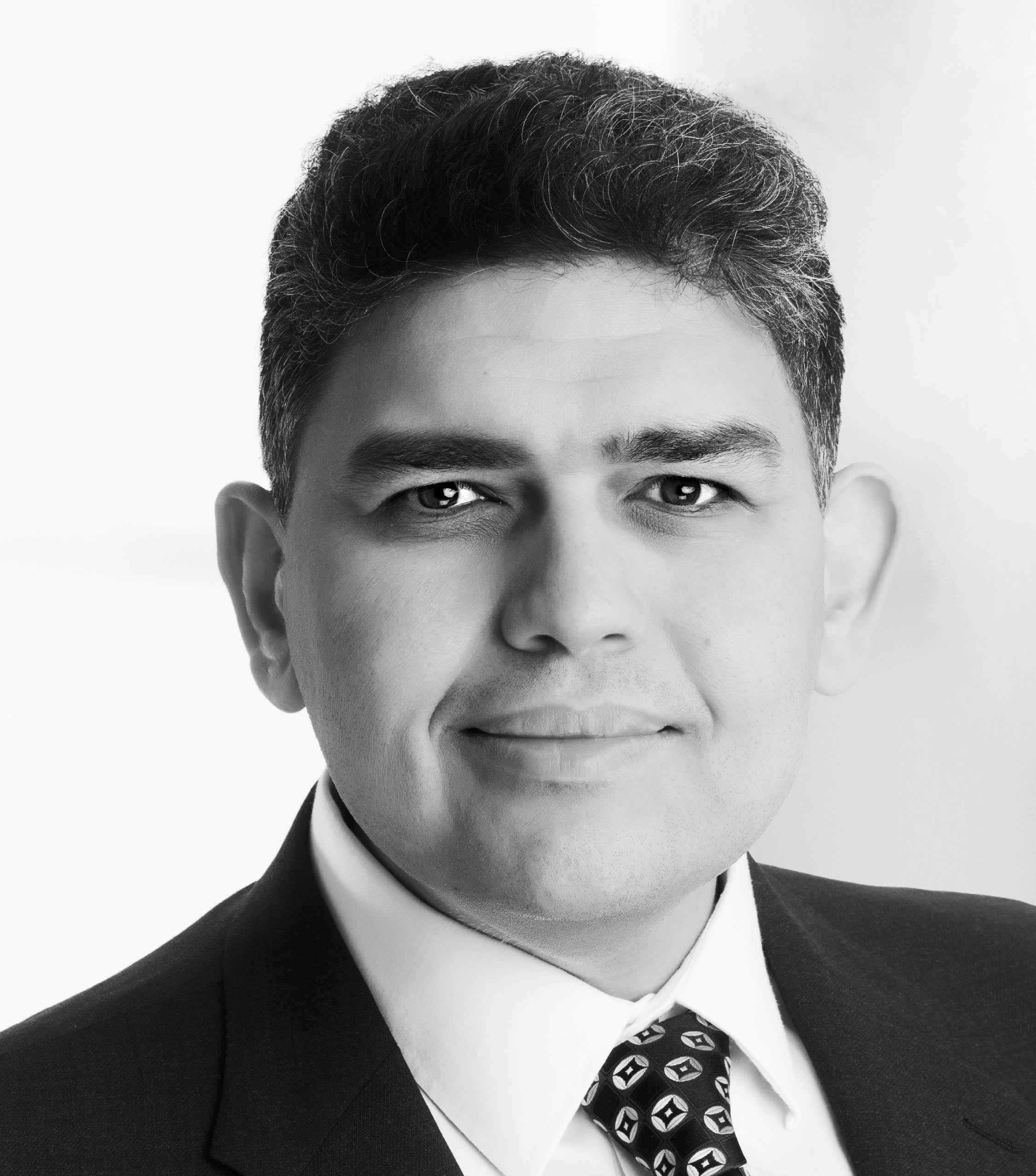}}]{Ahmed E. Hassan} is an IEEE Fellow, an ACM SIGSOFT Influential Educator, an NSERC Steacie Fellow, the Canada Research Chair (CRC) in Software Analytics, and the NSERC/BlackBerry Software Engineering Chair at the School of Computing at Queen’s University, Canada. His research interests include mining software repositories, empirical software engineering, load testing, and log mining. He received a PhD in Computer Science from the University of Waterloo. He spearheaded the creation of the Mining Software Repositories (MSR) conference and its research community. He also serves/d on the editorial boards of IEEE Transactions on Software Engineering, Springer Journal of Empirical Software Engineering, and PeerJ Computer Science. More information at: \url{http://sail.cs.queensu.ca}
\end{IEEEbiography}
\vspace{-0.9cm}
\begin{IEEEbiography}[{\includegraphics[width=1in,height=1.25in,clip,keepaspectratio]{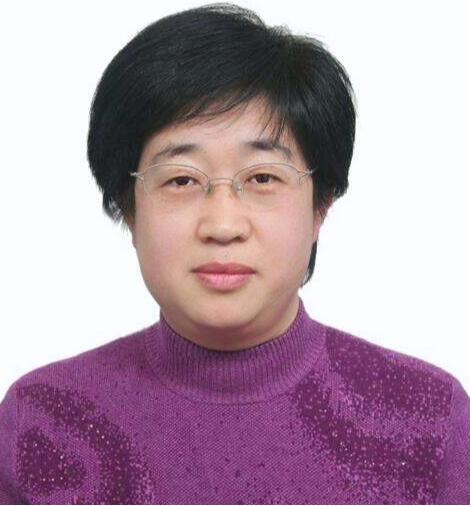}}]{Shujuan Jiang}
was born in Laiyang City, Shandong Province, in 1966. She received the B.S. degree in Computer Science of Computer Science Department from East China Normal University, Shanghai, in 1990 and the M.S. degree in Computer Science from China University of Mining and Technology, Xuzhou, Jiangsu, in 2000.  She received the Ph.D. degree in Computer Science from Southeast University, Nanjing, Jiangsu, in 2006.
From 1995 to today, she has been a teaching assistant, lecturer, associate professor and professor in the Computer Science Department, China University of Mining and Technology, Xuzhou, Jiangsu. She is the author of more than 80 articles. Her research interests include software engineering, program analysis and testing, software maintenance, etc. She is a Member of the IEEE.

\end{IEEEbiography}
\newpage


\end{document}